\documentclass[11pt,letterpaper]{article}
\usepackage{arxiv}
\usepackage[T1]{fontenc}
\usepackage[pdftex]{graphicx}
\usepackage{geometry}
\usepackage{rotate}
\usepackage{wrapfig}
\usepackage{caption}
\usepackage{subfigure}
\usepackage{times}
\usepackage{lineno,hyperref}
\usepackage{amsmath,amstext,amsbsy,amsopn,amsthm,amsfonts,amssymb}
\usepackage{url}
\usepackage{lastpage}
\usepackage{algorithm}
\usepackage{balance}
\usepackage{placeins}
\usepackage{subfigure}
\usepackage{caption}
\usepackage{capt-of}
\usepackage{algorithmic}
\usepackage{fancyhdr}
\usepackage{orcidlink}
\usepackage{wrapfig}
\providecommand{\keywords}[1]{\textbf{\textit{Keywords---}} #1}
\begin{document}

\title{iGLU 5.0: A Novel, Non-invasive and Intelligent HbA1c Measurement Device using Glucose values and Physiological Parameter for Smart Healthcare}

\author{
	\begin{tabular}{ccc}
		Prateek Jain \orcidlink{0000-0002-8191-9785} & Amit M. Joshi\orcidlink{0000-0001-7919-1652}& Saraju P. Mohanty \orcidlink{0000-0003-2959-6541} \\
		Electronics \& Instr. Eng. & Electronics \& Commu. Eng. & Computer Science and Engineering   \\
		Nirma University, India & 	MNIT, Jaipur, India. & University of North Texas, USA. \\
		prateek.jain@nirmauni.ac.in &amjoshi.ece@mnit.ac.in & smohanty@ieee.org
	\end{tabular}	
}

\maketitle

\cfoot{Page -- \thepage-of-\pageref{LastPage}}
\begin{abstract}
The laboratory test process of HbA1c measurement is a time-consuming and invasive method. The HbA1c parameter is the most important feature to predict the level of diabetes. Although invasive methods are irritating in the case of frequent measurements. Moreover, HbA1c measurement is only possible at the diagnostic centre, followed by medical protocol. Hence, it is still challenging to measure the HbA1c frequently at remote locations, where diagnostic centres are not easily available. Therefore, an intelligent and new non-invasive HbA1c measurement system, iGLU 5.0, is proposed for instant diagnosis of the HbA1c value without prior measurement setup. The proposed measurement device is based on optical spectroscopy for the collection of glucose values in different formats. The glucose values have been collected in fasting, postprandial, and random formats. The glucose value has also been collected using the oral glucose tolerance test (OGTT), along with the average blood pressure value, correspondingly. These four formats of glucose values, along with blood pressure, were used to predict the estimated average glucose (eAG) using an optimized prediction model. Further, the predicted average glucose is converted into an HbA1c value using a standard formula. The eAG prediction models have been trained and validated using 2000 samples of healthy, prediabetic and diabetic people to analyze the optimized prediction model. 94\% and 96\% accuracy have been examined during training and cross-validation of optimized DNN model, respectively. A 0.3 mean absolute difference has been identified from predicted HbA1c values using the proposed DNN model. The novel non-invasive HbA1c prediction system is useful for instant diagnosis without irritation for smart healthcare.
\end{abstract}

%
\section{Introduction}
Nowadays, diabetes is a critical issue for people around the world \cite{jainaccess}. It has become a primary disease due to an unhealthy lifestyle. People are following a prescribed diet and a prior treatment plan to mitigate the probability of being diabetic \cite{jain2024iglu}. Moreover, people are also looking towards the various advanced techniques and methods for early diabetes diagnosis \cite{hossain2021derivation}. There are several risk factors and corresponding tests for diagnosis to confirm diabetes \cite{jain2025}. HbA1c level is a major factor responsible for confirming diabetes \cite{bent2021non}. An HbA1c test is possible at the diagnostic centre, where a blood sample is taken from the human body for further processing. The laboratory test for HbA1c measurement is considered the gold standard from an accuracy point of view \cite{kaliappan2024analyzing}. The HbA1c level represents the three-month average glucose level for individuals. Based on the range of HbA1c values, people can be segregated as healthy or diabetic. Below the HbA1c level of 5.7, an individual is considered healthy. In the range of 5.7-6.4, the person is considered prediabetic. Beyond the HbA1c level of 6.4, diabetes will be confirmed. For the HbA1c test, a prior laboratory setup is required, and a significant amount of time is required for examination. Moreover, an invasive test is also not a favourable solution for frequent measurement. In rural areas and remote locations, there is less availability of pathology or laboratories for the HbA1c test, so people are not aware of the diseases or their early diagnosis. Hence, it is required to design a system which will not only confirm the level of diabetes on a prior basis but also provide the HbA1c value as well \cite{Jain2024_1}.

The overall paper has been organized as follows. Prior related work is reported in Section \ref{Sec:Prior-Works}. The domination of predictors and their role are demonstrated in Section \ref{Sec:parameters}. Novel research contributions of the proposed work are enlightened in Section \ref{Sec:Novel}. The proposed system of non-invasive HbA1c measurement is explained in detail with the process of HbA1c prediction in Section \ref{Sec:Proposed-Work}. Section \ref{Sec:Calibration} explored the methodology of prediction and computing models for system training and cross-validation. The experimental and measurement analysis has been done in Section \ref{Sec:error analysis}.

\section{Prior Related Research Work}
\label{Sec:Prior-Works}
An HbA1c measurement plays an important role in predicting the diabetes-level \cite{gomez2022understanding}. Graue et al. represented the primary role of HbA1c measurement to examine diabetes \cite{arnardottir2023using}. An analysis has been done on variations in diabetes diagnosis using fasting glucose and HbA1c \cite{ncd2023global}. Various methods are highlighted and utilized for HbA1c measurement. Mandali et al. explored the trends in estimation of HbA1c using electrochemical processes \cite{mandali2023trends}. They represented a cost-effective and point-of-care solution. In this way, Thapa et al. explained the glucose and HbA1c prediction using electrochemical detection \cite{thapa2023label}. Kwon et al. explored in vivo estimation of HbA1c using the PPG signal. They utilized machine learning models on PPG signal to extract the data for HbA1c estimation \cite{kwon2022machine}.  Hossain et al. validated a grey-box model to estimate in vivo HbA1c values \cite{hossain2021derivation}. They claimed a high accuracy in HbA1c estimation. Sridevi et al. explored the system for HbA1c estimation using a quantum machine learning approach. A PPG signal has been used to extract the features for the classification of HbA1c ranges \cite{sridevi2025noninvasive}. Mandal et al. explained the salient features of in vitro optical sensor design for HbA1c estimation \cite{mandal2018vitro}. An HbA1c measurement has been done in real time for physical validation. Based on experiments, the samples were segregated to identify the diabetic and non-diabetic people. 
The prior related works are represented with optimized approaches and better results \cite{agrawal2022machine}. They represented the scope of further advancement in technology. However, the presented works are either sensor development for HbA1c measurement or in vivo and in vitro testing using a real-time system with improved accuracy. Still, there is a challenge in estimating the HbA1c value without pricking blood directly from a human being. Various non-invasive works have been done for glucose prediction, which has been shown in Table \ref{prior_work}. 
\begin{table}[htbp]
	\caption{Non-invasive Measurement, Technology and Specifications}
	\label{prior_work}
	\centering
	\begin{tabular}{llllll}
		\hline
		\textbf{Works}&Mechanism&Sensing&Cost&Reliability&Application\\
		\textbf{}&&&&&\\
		\hline		 \hline
		Song, et al. \cite{Song2015}&IMPS+NIRS&Sensing+&Moderate&Moderate&Glucose\\
		&&Measurement&&&Measurement\\
		\hline
		Pai, et al.	\cite{Pai2018}&Photo-acoustic&Sensing+&High&High&Glucose\\
		&&Measurement&&&Measurement\\
		\hline
		Jain, et al.	\cite{jain2019precise}&mNIRS&Sensing&Low&Moderate&Glucose\\
		&&&&&Measurement\\
		\hline
		Jain, et al.	\cite{Jain_IEEE-MCE_2020-Jan_iGLU1}&mNIRS&Sensing &Low&High&Glucose\\
		&&+Trained Model&&&Measurement\\
		\hline
		Joshi, et al.	\cite{iGLU2}&mNIRS&Sensing &Low&High&Serum Glucose\\
		&&+Trained Model&&&Measurement\\
		\hline
		Joshi, et al. \cite{iGLU3}&mNIRS+&Sensing &Low&High&Body Glucose, Insulin\\
		&Data Security&+Trained Model&&&Measurement\\
		\hline
		Jain, et al.\cite{9221132}&Mathematical&Trained&Moderate&Moderate&Glucose\\
		&Model&Model&&&Insulin Model\\
		\hline
			Murad, et al.\cite{new9431682}&mNIRS&Simulation&-&-&Glucose Detection\\
			\hline
			Kirubakaran, et al.\cite{newkirubakaran2023antiallergic}&Microwave&Sensing&High&Moderate&Glucose\\
			&&&&&Measurement\\
			\hline
			Mohammadi, et al.\cite{newmohammadi2023dual}&Microwave&Sensing&High&Moderate&Glucose\\
			&Resonator&&&&Detection\\
		\hline
		(iGLU 4.0)&mNIRS+&Sensing+&Moderate&High&Glucose Balancing\\
		\cite{jain2024iglu}&Model&Measurement&&&Paradigm\\
		\hline
        \textbf{(iGLU 5.0)}&\textbf{Glucose+}&\textbf{Sensing+}&\textbf{Moderate}&\textbf{High}&\textbf{HbA1c Measurement}\\
		&\textbf{BP}&\textbf{Measurement}&&&\textbf{}\\
		\hline
	\end{tabular}
\end{table}

\begin{figure}
	\centering
	\includegraphics[width=0.8\textwidth]{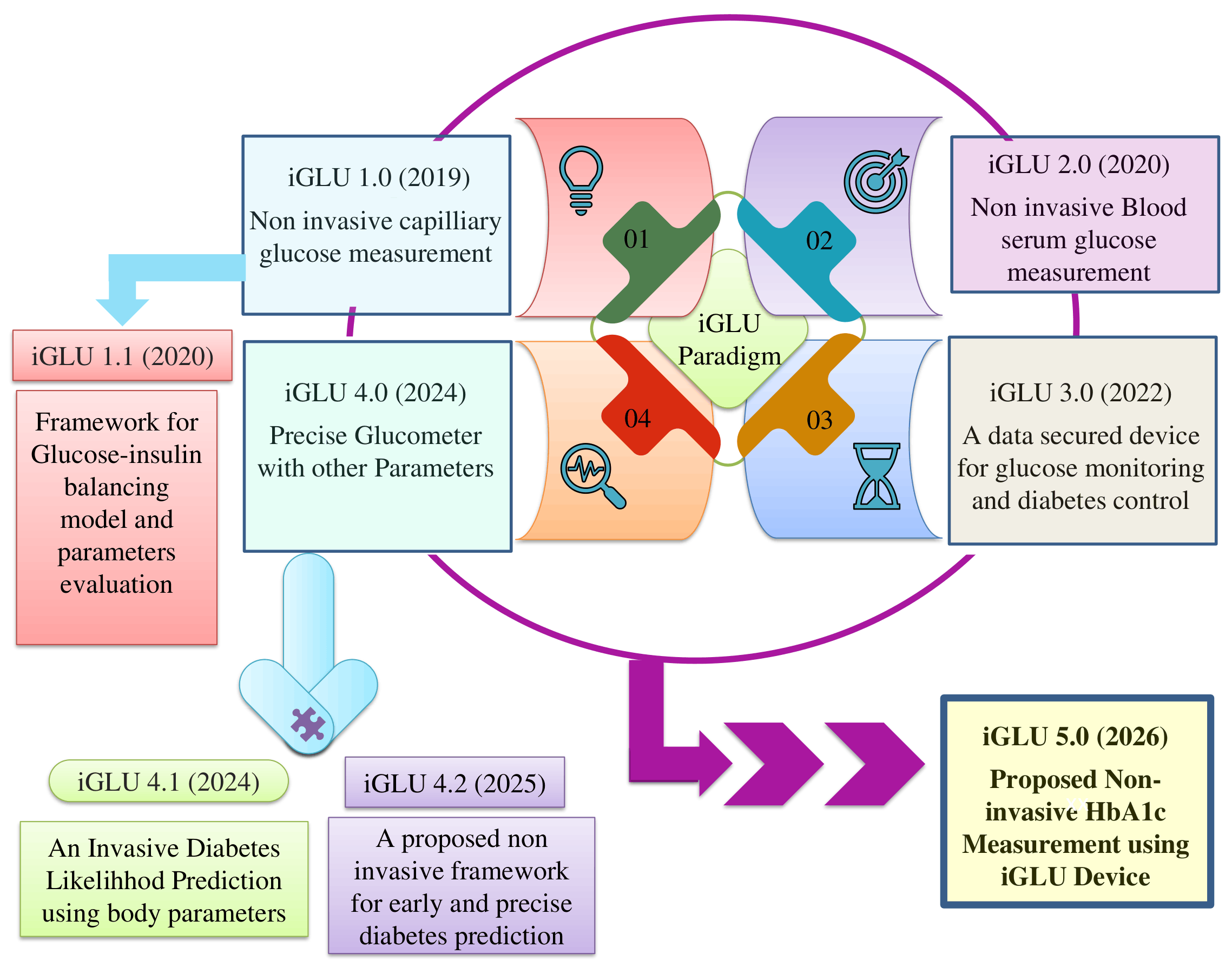}
	\caption{Evolution of iGLU with Applications}
	\label{prior}
\end{figure}

The prior works motivate for non-invasive HbA1c prediction. Hence, a non-invasive measurement is required, which can measure HbA1c without a prior setup for instant diagnosis. For this objective, a wearable, non-invasive HbA1c measurement device for smart healthcare. The proposed device is a precise, portable, and cost-effective solution for diabetes care. The proposed device, iGLU 5.0, is an advanced version of the iGLU device. This is demonstrated by different versions of the iGLU device. The technology, features, and applications are represented in Table \ref{prior_work1}.
\begin{table}[htbp]
	\caption{Different versions of iGLU  and Specifications}
	\label{prior_work1}
	\centering
	\begin{tabular}{llllll}
		\hline
		\textbf{Works}&Technology&Measurement&Predictors&Framework&Application\\
		\hline		 \hline
		iGLU 1.0	\cite{Jain_IEEE-MCE_2020-Jan_iGLU1}&Optical&Non-invasive &Light of&Hardware&Glucose\\
		&&&Specific &&Measurement\\
        &&& wavelength&&\\
		\hline
        iGLU 1.1 \cite{9221132}&Mathematical&Semi-&Glucose&Software&Glucose\\
		&Model&invasive&Profile&&Insulin Model\\
		\hline
		iGLU 2.0	\cite{iGLU2}&Optical&Non-invasive &Light of&Hardware&Serum Glucose\\
		&&&Specific &&Measurement\\
        &&& wavelength&&\\
		\hline
		iGLU 3.0 \cite{iGLU3}&mNIRS+&Non- &Lights of&Hardware+&Body Glucose, Insulin\\
		&Data Security&invasive&wavelength&Software&Measurement\\
		\hline
		(iGLU 4.0)&mNIRS+&Non-&Physiological&Hardware&Glucose Balancing\\
		\cite{jain2024iglu}&Model&invasive&Parameters&&Paradigm\\
		
        \hline
		(iGLU 4.1)&mNIRS+&Semi-&Physiological&Hardware+&Diabetes\\
		\cite{Jain2024_1}&Model&invasive&Parameters+BG&Software&Prediction\\
        \hline
		(iGLU 4.2)&mNIRS+&Non-&Physiological&Hardware+&Diabetes Likelihood\\
		\cite{jain2025}&Model&invasive&Parameters+BG&Software&Prediction\\
        \hline
        \textbf{(iGLU 5.0)}&\textbf{mNIRS+}&\textbf{Non-}&\textbf{Glucose}&\textbf{Hardware}&\textbf{HbA1c}\\
&\textbf{model}&\textbf{invasive}&\textbf{Values+BP}&&\textbf{Measurement}\\
		\hline
	\end{tabular}
\end{table}
This prior related work represents the motivation to upgrade an edge computing device, iGLU, for a specific application. The evolution of iGLU is represented in Fig. \ref{prior}. 
The proposed work is encouraged to upgrade from earlier versions of iGLU \cite{joshi2020smart}. The first version, iGLU 1.0, measures the capillary glucose from a boneless part of the human body. Its sub-version iGLU 1.1 represents the framework of the glucose-insulin model for the glucose-balancing paradigm. IGLU 2.0 explores the wearable solution for blood serum glucose measurement. iGLU 3.0 is designed for a secure device of glucose-insulin control. iGLU 4.0 explores the physiological parameters for precise blood glucose measurement. The proposed work, iGLU 5.0, uses the iGLU for collecting the glucose values of all formats (fasting, postprandial, random, and OGTT) to predict the HbA1c values directly from the human body.

\section{Significance of Predictors Addressed for HbA1c Measurement}
\label{Sec:parameters}

\begin{figure}
	\centering
	\includegraphics[width=1.0\textwidth]{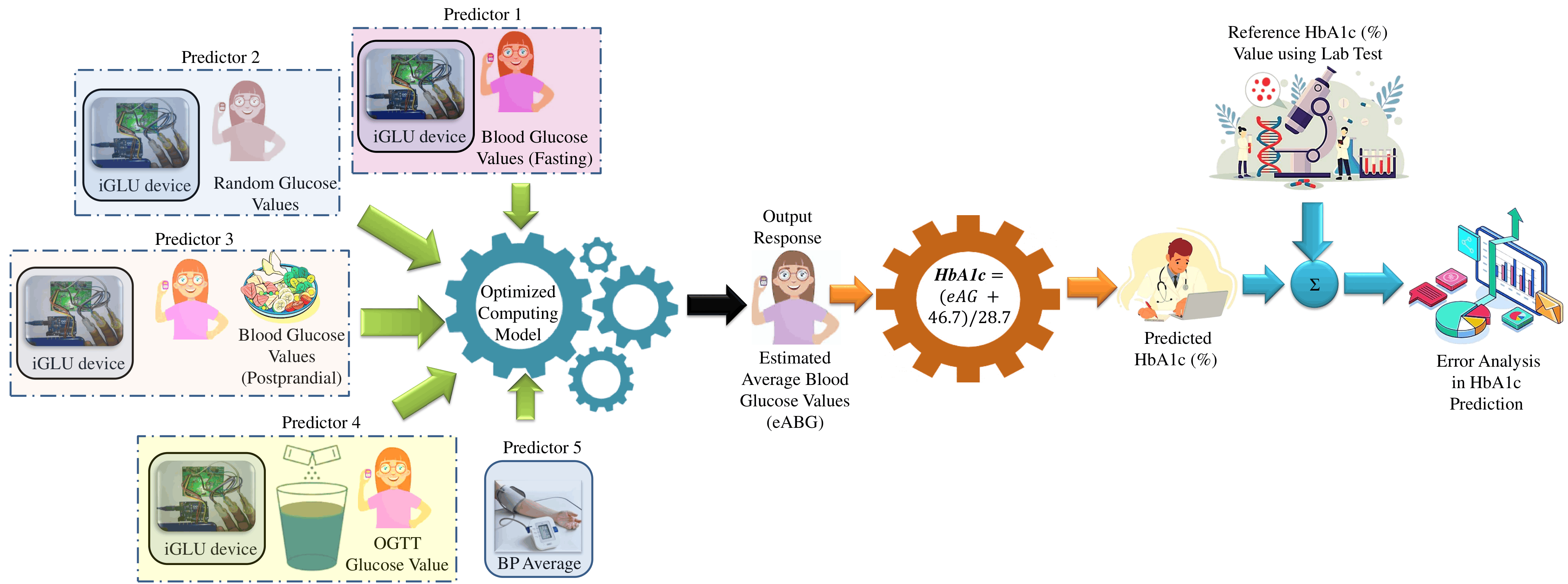}
	\caption{Overall Process Representation of HbA1c Measurement using iGLU 5.0}
	\label{process}
\end{figure}

Accurate diabetes prediction is challenging as it depends on various factors \cite{MCE}. HbA1c values are most responsible factor for proper diagnosis. Estimation of average glucose is directly related to the HbA1c value. Generally, three months average glucose values are being consider for HbA1c prediction \cite{gough2023within}. The glucose values in four different formats (fasting, postprandial, random, and OGTT) are significantly analyzed for precise prediction of the estimated average glucose value. OGTT refers to oral glucose tolerance test. For OGTT glucose value, the person has to take 75g of glucose after fasting of 8 hours \cite{Liu2016}. After 1 hour of 75g oral glucose intake, the glucose value is measured for a precise OGTT value \cite{beunen2022fasting}. For random glucose, the glucose value is being taken throughout the day. Fasting glucose is measured after 8-12 hours since the last meal \cite{jamieson2023oral}. Postprandial glucose value is being measured after 2-3 hours of meal intake. Since glucose molecules are insoluble in blood, they are distributed in the blood asynchronously. Therefore, blood pressure plays an important role in precise measurement \cite{jain2024iglu}. These four glucose values and average blood pressure are appropriate to predict the precise average glucose value for HbA1c prediction. These five predictors are used as independent variables to examine the computing model for eAG prediction \cite{dahal2023predicting}. After analyzing the optimized model, the precise eAG value has been predicted to evaluate the HbA1c value. The complete process of HbA1c prediction has been visualized in Fig \ref{process}.
The proposed system is trained to predict a precise eAG value using the five mentioned predictors (inputs). Then, the system is tested through comparison of non-invasive and corresponding invasive predicted HbA1c values for error analysis.


\section{Novel Contribution of Proposed Work}
\label{Sec:Novel}

The proposed HbA1c measurement system is painless, which provides the exact measurement of the average glucose value for instant treatment. It also minimizes the chances of having blood-related diseases as the non-invasive measurement is being done. The optical approach of the iGLU device, along with optimized prediction model of eAG prediction, has provided an optimal solution for non-invasive HbA1c measurement. The proposed iGLU 5.0 resolved the issue of instant diagnosis at remote or rural locations, where a laboratory setup may not be available. The proposed system does not need prior setup for measurement. It can be used to measure the HbA1c value without prior recommendation of medical expert. The system will be portable for frequent use anytime and anywhere. The system is an easy-to-handle, rapid and low-cost solution for smart healthcare.
The proposed iGLU 5.0 framework is also the reliable solution at the patient care centre for rapid measurement without prior setup. It will be a useful setup for medical experts to provide prescribed treatment to the patients.

The \textbf{novel contribution in proposed work} are as follows: 
\begin{enumerate}
	\item 
	An optimized and accurate non-invasive HbA1c measurement system has been developed, which would be able to provide the diagnosis solution at the remote locations. 
	\item 	An optimized HbA1c prediction model has been analyzed, and trained from collected glucose and BP values of 2000 samples of balanced healthy and diabetic people. 
	\item The optimized DNN model has been cross-validated and tested using different healthy and diabetic people for an accurate HbA1c diagnosis system.
 
\end{enumerate}
\section{Proposed Non-invasive HbA1c Measurement System iGLU 5.0}
\label{Sec:Proposed-Work}
\begin{figure}[htbp]
	\centering
	\includegraphics[width=1.0\textwidth]{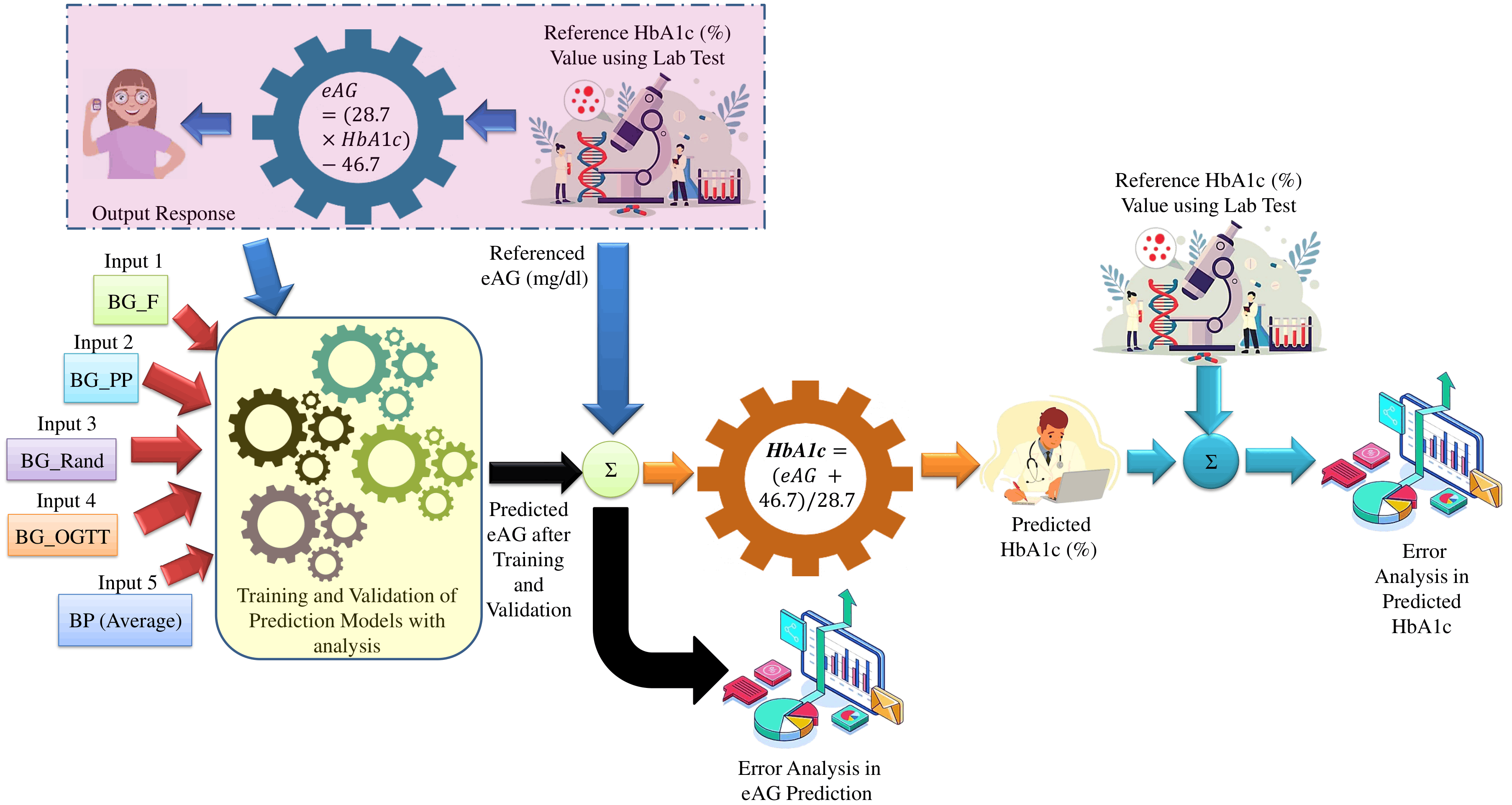}
	\caption{Proposed Methodology for Non-invasive HbA1c Measurement}
	\label{methodology}
\end{figure}

The proposed HbA1c measurement device is based on predictors (glucose values in different formats and average blood pressure). The glucose values are measured using a non-invasive mechanism. Average blood pressure values have also been logged from individuals. The logged predictors are non-invasive values. The glucose values are collected using optical spectroscopy through specific wavelengths of glucose molecule detection. All predictor values provided estimated average glucose values using analyzed and optimized DNN model. The proposed methodology has been visualized in Fig. \ref{methodology}.
The different glucose values with average blood pressure have been chosen for eAG prediction. Initially, conventional HbA1c values have been taken from individuals and converted into eAG using standard formula. At the same time, glucose and average blood pressure values have also been taken for system calibration. The set of predictor and corresponding eAG values (eAG obtained from conventional HbA1c test) are used to train the different prediction models. The optimized model has been recognized through error analysis of eAG predictions during training and validation. After the confirmation of optimized model, eAG is predicted without pricking blood. HbA1c values have also been obtained using standard formula, and error analysis has also been done in HbA1c prediction.
\section{Proposed Computing model of estimated average glucose for HbA1c Measurement}

\label{Sec:Calibration}

A post-processing computing model has been analyzed during training and validation of the proposed system. Different prediction models are tested for an accuracy point of view, as it is important to examine the post-processing model for precise HbA1c estimation \cite{Sarangi2014}. The data has been collected from healthy and diabetic people aged 18-75 years. The baseline characteristics of healthy and diabetic people are represented in Table \ref{dataset}.

\begin{figure}[htbp]
	\centering
	\includegraphics[width=1.0\textwidth]{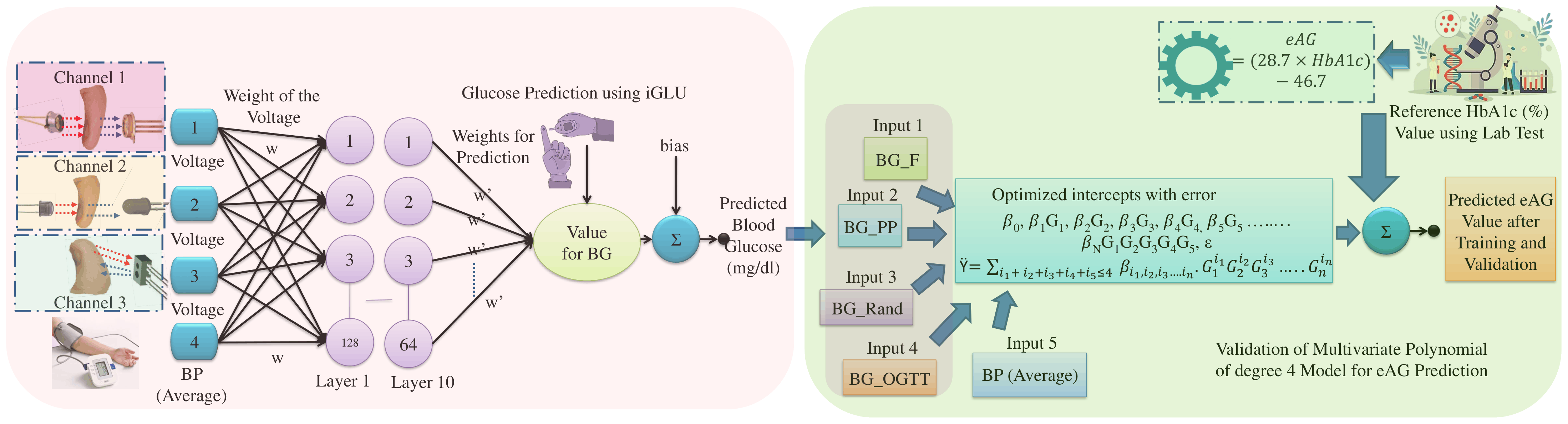}
	\caption{Computational Processing using MVPD4 for eAG Prediction}
	\label{Schematic}
\end{figure}

\begin{table}[htbp]
	\caption{Data Summary of Collected Samples for HbA1c Prediction}
	\label{dataset}
	\centering
	\begin{tabular}{lll}
		\hline
		\hline
		\textbf{Basic} & \textbf{Subjects for} & \textbf{Subjects for}\\
		\textbf{Attributes} & \textbf{Training} & \textbf{ Validation and Testing}\\
		\hline
		\textbf{Gender} & \textbf{Gender wise} & \textbf{Gender wise}\\
		Male:-   \textbf{1200} & Male:-   \textbf{800} & Male:-   \textbf{400}\\
		Female:- \textbf{800} & Female:- \textbf{600} & Female:- \textbf{200}\\
		\hline
		\textbf{Age (Years)} & \textbf{Age wise} & \textbf{Age wise}\\
		Male:-   \textbf{22-65} & 18-47 Yrs:-   \textbf{375}& Not segregated\\
		Female:- \textbf{18-75} & 48-75 Yrs:- \textbf{1025} & \\
		\hline
		\hline
	\end{tabular}
\end{table}
Various prediction models are trained to analyze an optimized computing model for real-time eAG measurement. Two different models, the multivariate polynomial model with degree n (MVPDn) and deep neural network (DNN), are observed for precise measurement.
To train the MVPDn model, the predictors are considered as $BG_F$, $BG_{PP}$, $BG_R$, $BG_O$, and $BP_A$, and the output is $eAG$.
The predictors is represented in vector form by eq. \ref{1}.
\begin{equation}
\mathbf{G}= \begin{bmatrix} BG_F \\ BG_{PP} \\ BG_R \\ BG_O \\ BP_A \end{bmatrix} 
\label{1}
\end{equation}
where, $\mathbf{G} \in \mathbb{R}^{5}$, Then the computing model can be attempted to approximate
\begin{equation}
eAG=f(BG_F, BG_{PP}, BG_R, BG_O, BP_A)+\epsilon 
\label{2}
\end{equation}
Here, f(.)f(.)f(.) is	polynomial function, and
$\epsilon$	is a random error term.
After analyzing the polynomial function, the degree 4 has been observed as a precise model.
A degree 4 polynomial function includes all terms:\\
$i_1+i_2+i_3+i_4+i_5\le4$\\
The general polynomial model can be expressed as
\begin{equation}
eAG= \sum_{|\alpha|\le4} \beta_{\alpha}\mathbf{G}^{\alpha} +\epsilon 
\label{3}
\end{equation}
Here, $\alpha= (\alpha_1,\alpha_2,\alpha_3,\alpha_4,\alpha_5)$, and $\alpha_i\ge0$.
The total degree of multi-index vector can be represented as 
\begin{equation}
|\alpha|= \alpha_1+\alpha_2+\alpha_3+\alpha_4+\alpha_5  \nonumber
\end{equation}
Hence, polynomial basis term can be represented in eq. \ref{4},
\begin{equation}
\mathbf{G}^{\alpha} = BG_F^{\alpha_1} BG_{PP}^{\alpha_2} BG_R^{\alpha_3} BG_O^{\alpha_4} BP_A^{\alpha_5} 
\label{4}
\end{equation}
Here, the total predictors are 5, and degree of polynomial is 4. Then, total polynomial terms are
$T= \binom{p+d}{d}$.\\
Substitute the values of $p$ and $d$, then
$T= \binom{5+4}{4}$.
Hence, total polynomial terms including intercepts are 126.
The prediction equation can be represented in eq. \ref{5}.
\begin{eqnarray}
\hat{eAG} = [\sum_{|\alpha|\le4} \beta_{\alpha} BG_F^{\alpha_1} BG_{PP}^{\alpha_2} BG_R^{\alpha_3} BG_O^{\alpha_4} BP_A^{\alpha_5}]+\epsilon
\label{5}
\end{eqnarray}
The diagram of computational processing using MVPD4 for eAG prediction is given in Fig. \ref{Schematic}.

\begin{figure}[htbp]
	\centering
	\includegraphics[width=1.0\textwidth]{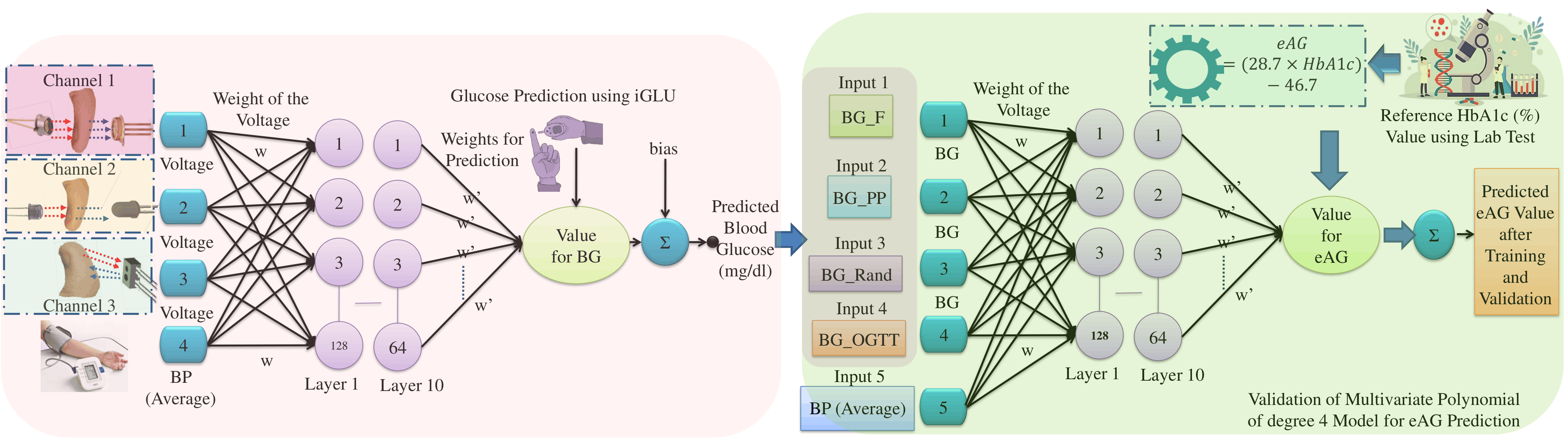}
	\caption{Computational Processing using DNN model for eAG Prediction}
	\label{Schematic1}
\end{figure}

DNN based model has also been analyzed for precise eAG prediction after training and cross-validation.
For the computing model, customized weights are connected with neurons between the hidden layers.
It can be expressed in eq. \ref{6}.
\begin{equation}
W^{(1)}\in\mathbb{R}^{16\times5}  
\label{6}
\end{equation}
Here, 16 neurons are in hidden layers. Eq. \ref{6} can be expanded and represented in eq. \ref{7}.
\begin{equation}
W^{(1)}= \begin{bmatrix} w_{11}^{(1)} & w_{12}^{(1)} & \cdots & w_{15}^{(1)}\\ w_{21}^{(1)} & w_{22}^{(1)} & \cdots & w_{25}^{(1)}\\ \vdots & \vdots & \ddots & \vdots\\ w_{16,1}^{(1)} & w_{16,2}^{(1)} & \cdots & w_{16,5}^{(1)} \end{bmatrix}  
\label{7}
\end{equation}
After the summation of weights of hidden layers, the linear transformation can be expressed as
\begin{equation}
p^{z(l)}=W^{(l)}A^{(l-1)}+b^{(l)} 
\label{8}
\end{equation}
Hence, complete DNN function can be represented after deploying sigmoid activation for prediction.
\begin{equation}
\hat{eAG} = W^{(11)} T\left( W^{(10)} T\left( \cdots T\left( W^{(1)}p+b^{(1)} \right) \right) \right) +b^{(11)}
\label{9}
\end{equation}
The final computing model for eAG prediction is represented by eq. \ref{10}.
\begin{equation}
\hat{eAG}=W^{(11)}A^{(10)}+b^{(11)}
\label{10}
\end{equation}
The diagram of computational processing using optimized DNN for eAG prediction is given in Fig. \ref{Schematic1}.

After precise eAG prediction through an optimized model, the HbA1c can be predicted using a mathematical expression, which is presented in eq. \ref{11}.
\begin{equation}
HbA1c (\%)=(eAG+46.7)/28.7
\label{11}
\end{equation}
\section{Comparative Analysis of Experimental Results}
\label{Sec:error analysis}

The presented computing model used five predictors for average glucose prediction. The data has been collected as per the prescribed standards and medical protocols. The balanced data has been prepared from healthy, pre-diabetic and diabetic patients. The data has also been considered as age and gender-wise, which is shown in Table \ref{dataset}. Experimental analysis of performance parameters has been done among the calibrated models for precise eAG measurement. For calibration of computing models, MATLAB 2026a version has been utilized on an Intel Core i7, $10^{th}$ generation processor system with a 256 GB SSD for fast processing. 40 seconds of processing time has been identified for DNN computing model calibration. An 800 obs/sec computation speed has also been confirmed during calibration and validation time. The optimized DNN-trained model size has been confirmed as a 15.2 KB file size. Different prediction models have been trained,and DNN model has been justified as an optimized model. The comparative analysis of accuracy parameters is mentioned in Table \ref{2t} and \ref{2t1}.
\begin{table}[htbp]
\caption{Comparative Analysis of different Computing Models During Training}
\label{2t}
\centering
\begin{tabular}{lccc}
\hline
\multicolumn{1}{l}{\textbf{Models}} & \multicolumn{1}{l}{\textbf{MAD (HbA1c)}} & \multicolumn{1}{l}{\textbf{mARD}} & \textbf{AvgE (\%)}  \\ \hline
MVPD2                          & 0.49 &0.08  & 7.91               
\\ \hline
MVPD3                     & 0.39 &0.062  & 6.045               
\\ \hline
MVPD4                        & 0.38 &0.061  & 5.96               
\\ \hline
Fine-Gaussian SVR                 & 0.47 &  0.08& 7.0               
\\ 
 \hline
DNN                 & 0.39 &0.061  & 6.17 
\\
\hline
\end{tabular}
\end{table}

\begin{table}[htbp]
\caption{Comparative Analysis of different Computing Models During Validation and Testing}
\label{2t1}
\centering
\begin{tabular}{lccc}
\hline
\multicolumn{1}{l}{\textbf{Models}} & \multicolumn{1}{l}{\textbf{MAD (HbA1c)}} & \multicolumn{1}{l}{\textbf{mARD}} & \textbf{AvgE (\%)}  \\ \hline
MVPD2                          & 0.41 &0.05  & 5.25               
\\ \hline
MVPD3                     & 0.38 &0.048  & 4.75               
\\ \hline
MVPD4                        & 0.377 &0.048  & 4.707               
\\ \hline
Fine-Gaussian SVR                 & 0.43 &  0.07& 6.0               
\\ 
 \hline
DNN                 & 0.376 &0.048  & 4.67 
\\
\hline
\end{tabular}
\end{table}

\begin{figure}[htbp]
	\centering
\includegraphics[width=0.9\textwidth]{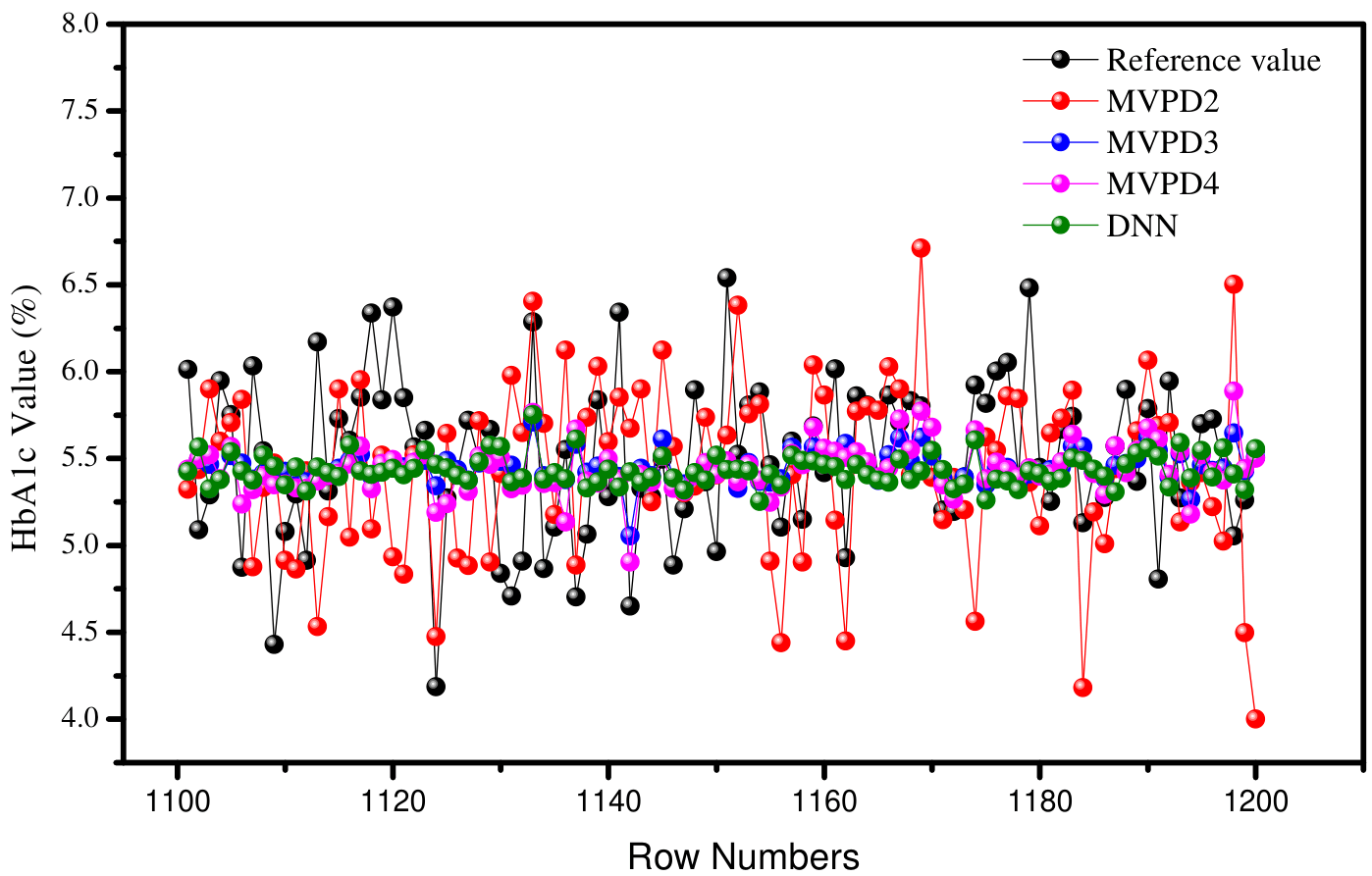}
	\caption{HbA1c predicted values of healthy people using the conventional method and computing models}
	\label{healthy}
\end{figure}

\begin{figure}[htbp]
	\centering
\includegraphics[width=0.9\textwidth]{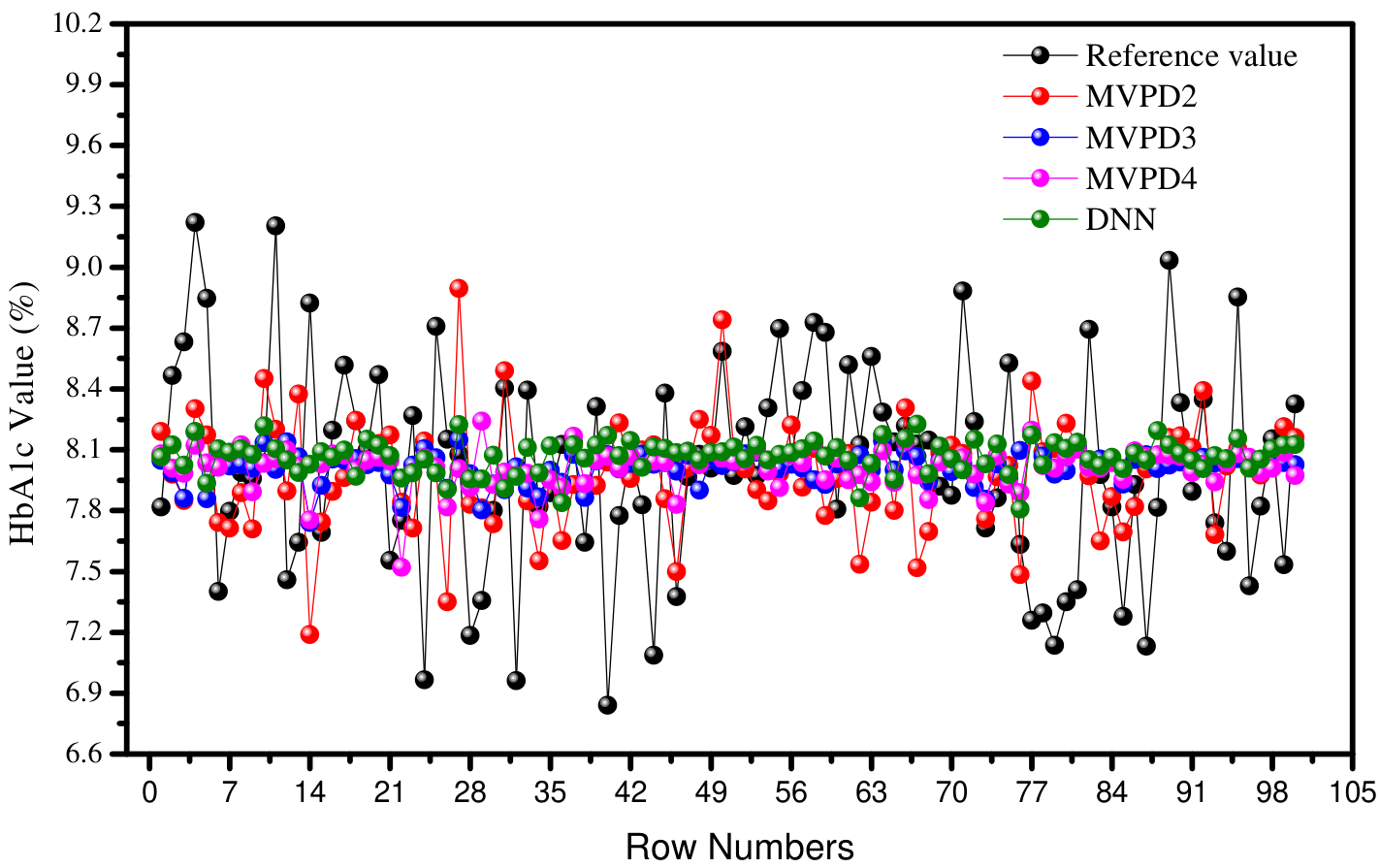}
	\caption{HbA1c predicted values of hyperglycemic patients using conventional method and computing models}
	\label{diabetic}
\end{figure}

\begin{figure}[htbp]
	\centering
\includegraphics[width=0.7\textwidth]{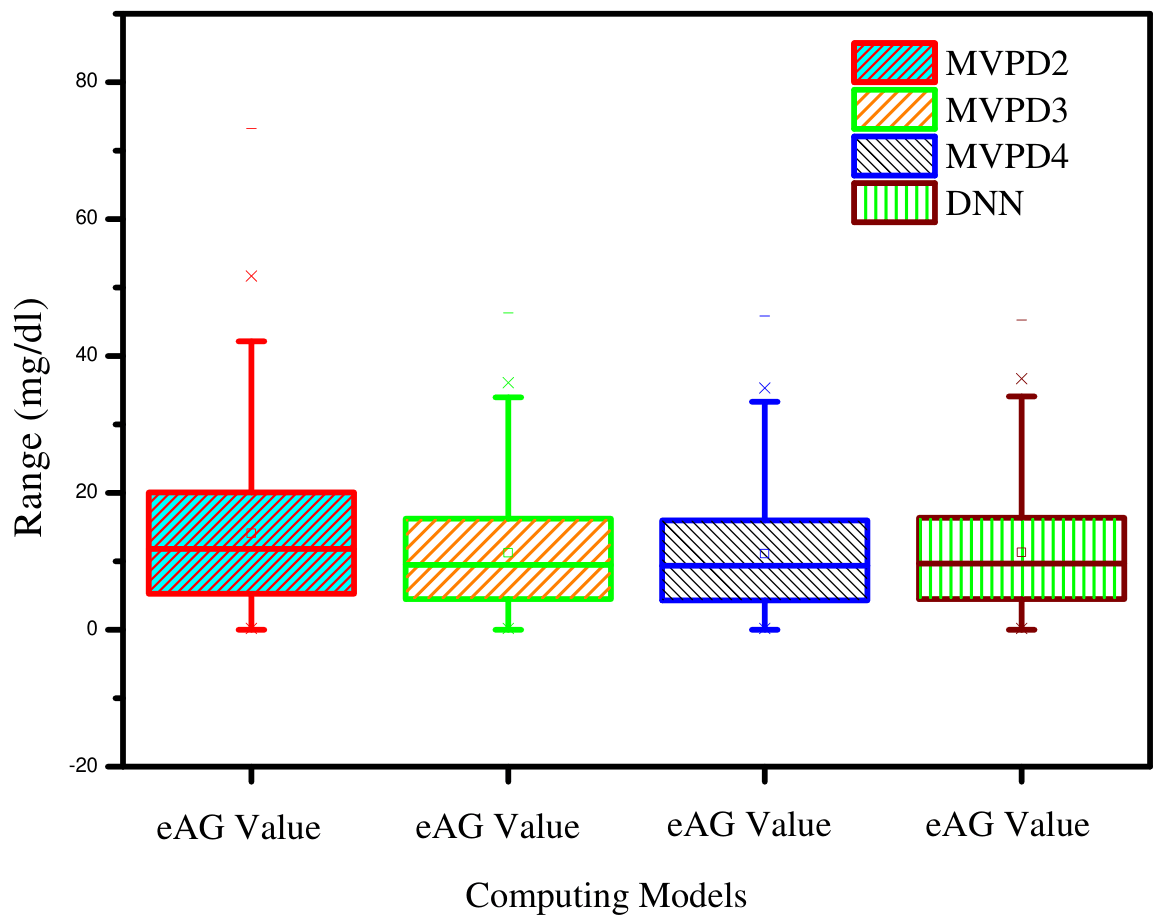}
	\caption{Error range representation in eAG prediction}
	\label{box1}
\end{figure}

\begin{figure}[htbp]
	\centering
\includegraphics[width=0.7\textwidth]{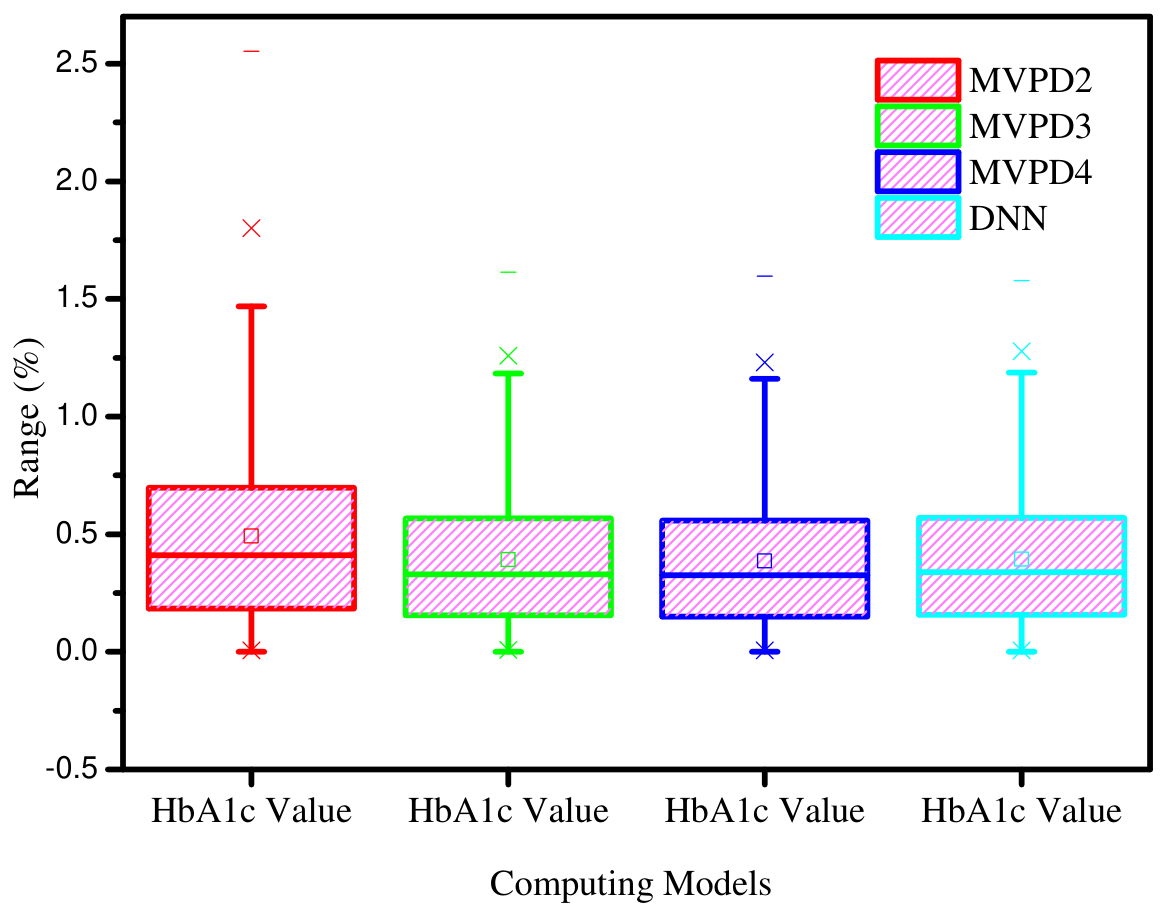}
	\caption{Error range representation in HbA1c prediction}
	\label{box2}
\end{figure}

For accuracy analysis of computing models, a 2000-sample of diabetic and healthy people have been taken for calibration, validation, and testing. 70\% of balanced (healthy and diabetic) samples have been randomly chosen for model calibration, and 30\% has been taken for validation and testing. 5-fold cross-validation has been done after calibration of different models for accuracy analysis. 200 different samples have been bifurcated to cross-test the computing models. 
After the quantitative analysis, it has been concluded that 94\% and 95.5\% accuracy have been confirmed using optimized DNN model, respectively.
Randomly selected predicted HbA1c responses using optimized models have been visualized in Fig. \ref{healthy} and \ref{diabetic}.
During calibration, validation, and testing of models, error analysis has been done for eAG and HbA1c prediction. The Error ranges have been evaluated in eAG and HbA1c prediction for different models.
The box plots are represented for eAG and HbA1c prediction in Fig. \ref{box1} and \ref{box2}.

\begin{figure}[htbp]
	\centering
\includegraphics[width=0.7\textwidth]{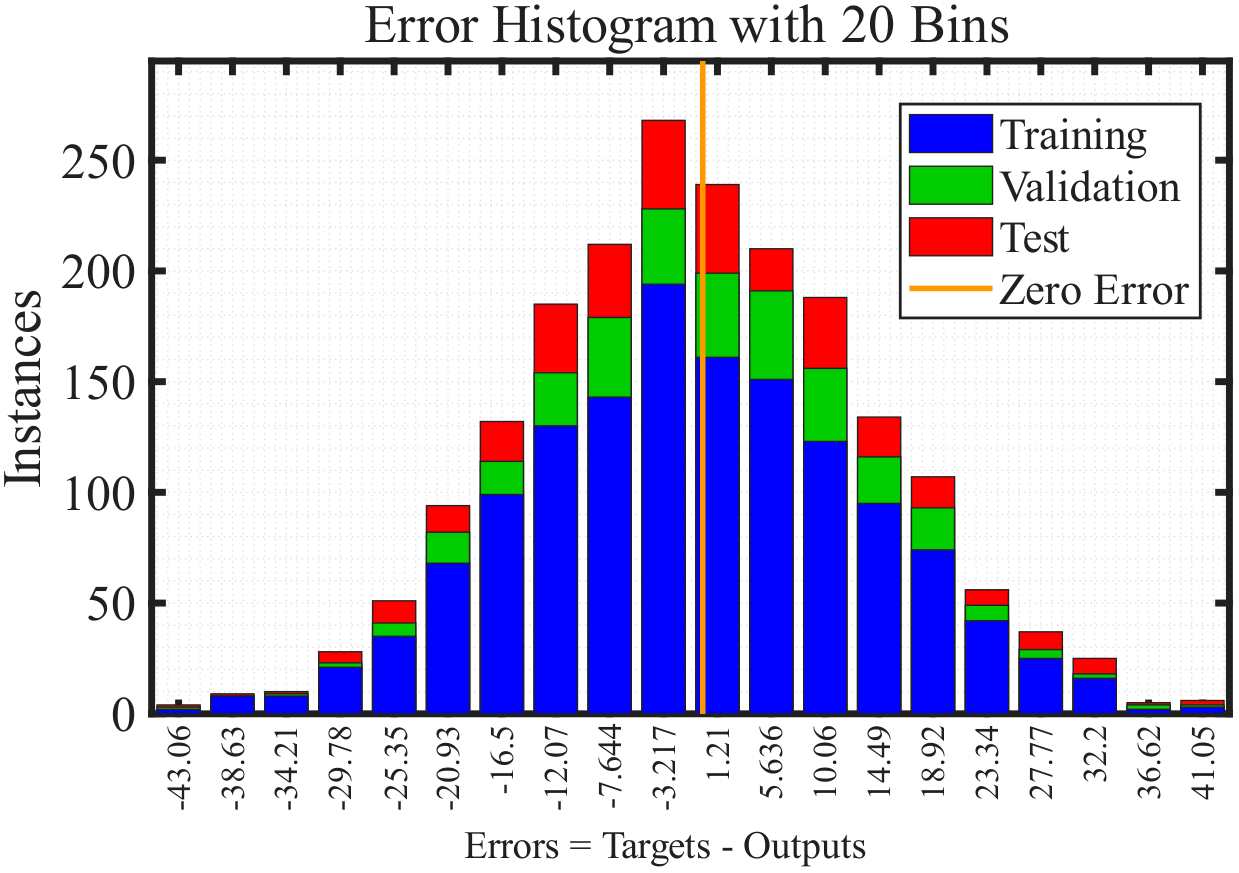}
	\caption{Error histogram representation with 20 bins in eAG prediction}
	\label{hist}
\end{figure}

\begin{figure}[htbp]
	\centering
\includegraphics[width=0.7\textwidth]{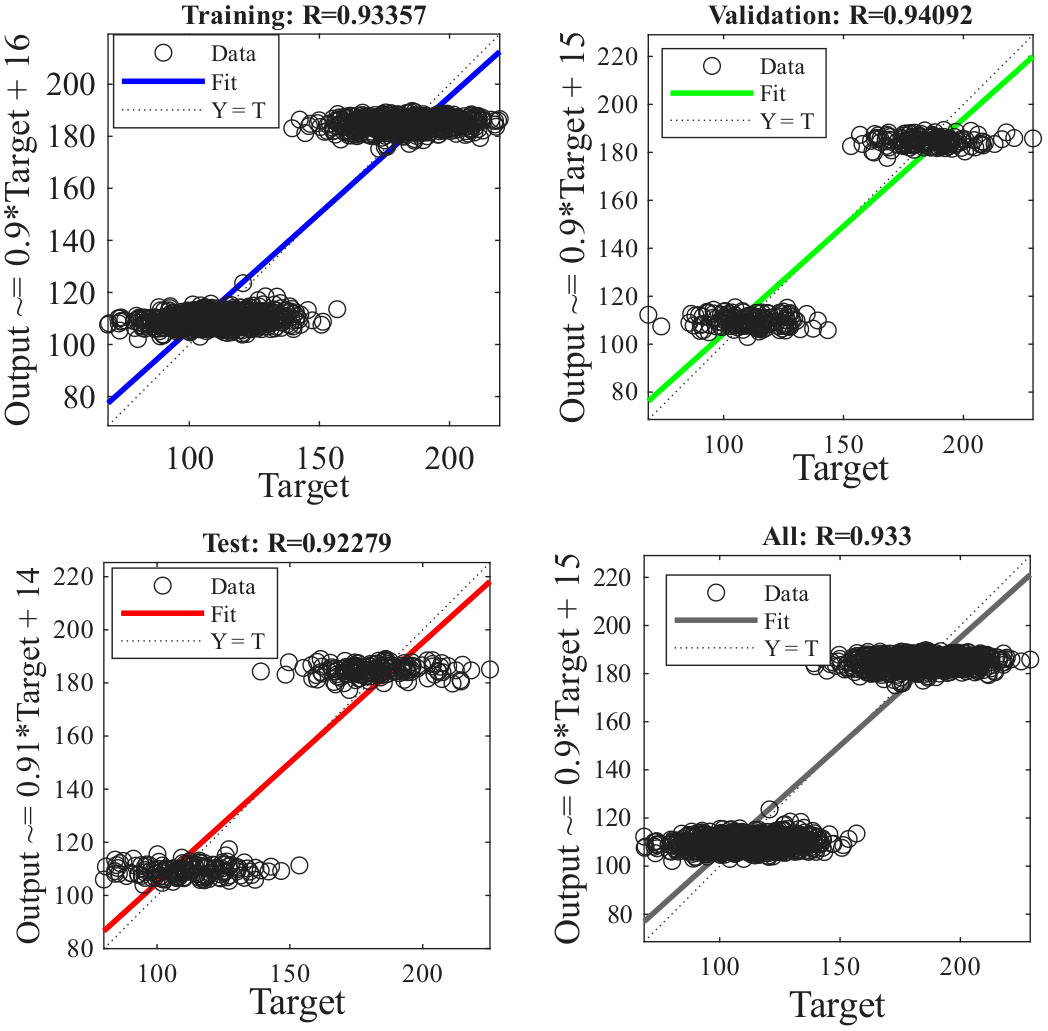}
	\caption{Correlation representation during calibration and testing using DNN model}
	\label{nnfit}
\end{figure}

\begin{figure}[htbp]
	\centering
	\includegraphics[width=0.6\textwidth]{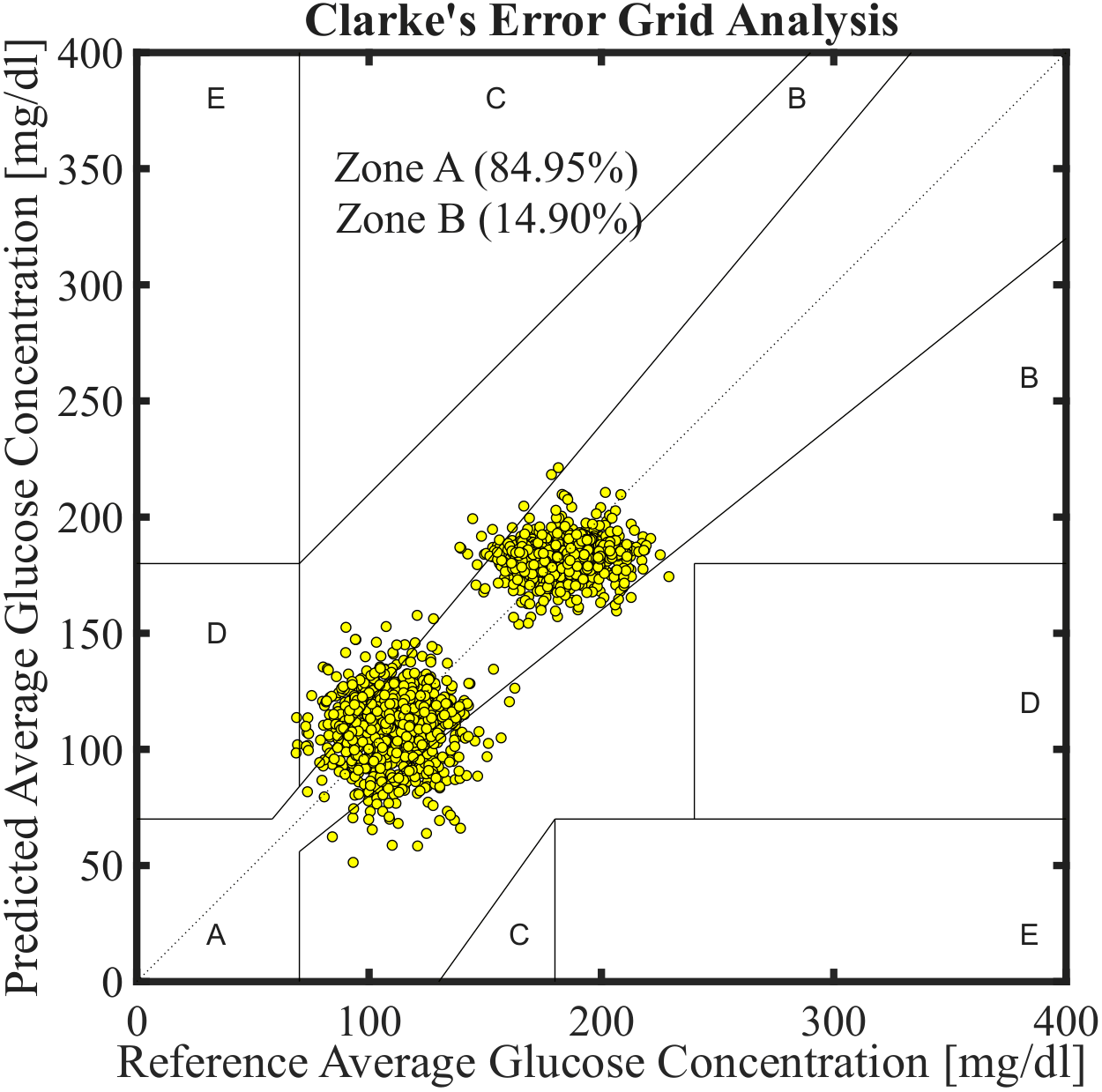}
	\caption{CEG analysis of training samples using MVPD2 model}
	\label{C2}
\end{figure}

\begin{figure}[htbp]
	\centering
	\includegraphics[width=0.6\textwidth]{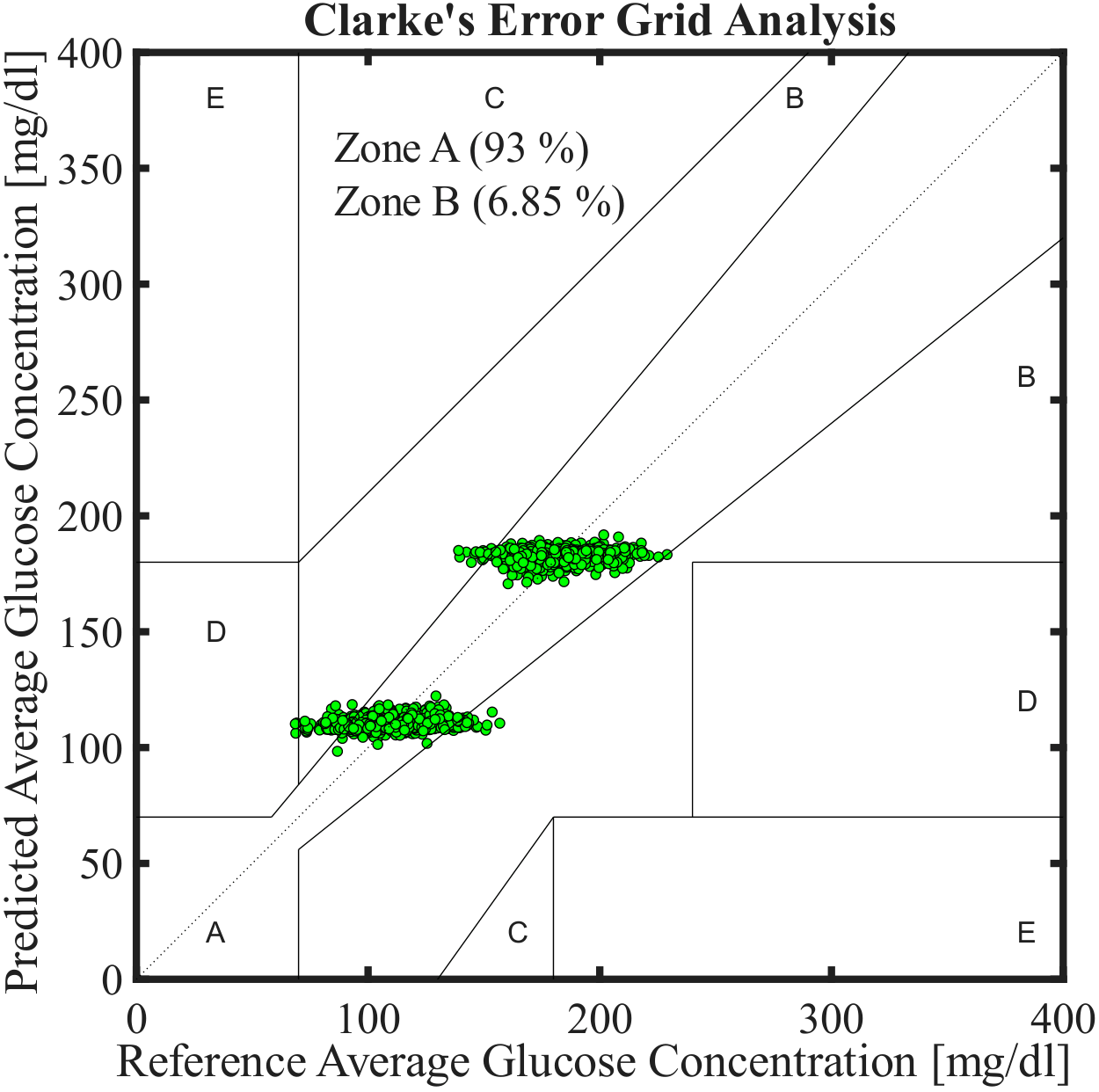}
	\caption{CEG analysis of training samples using MVPD3 model}
	\label{C3}
\end{figure}

\begin{figure}[htbp]
	\centering
	\includegraphics[width=0.6\textwidth]{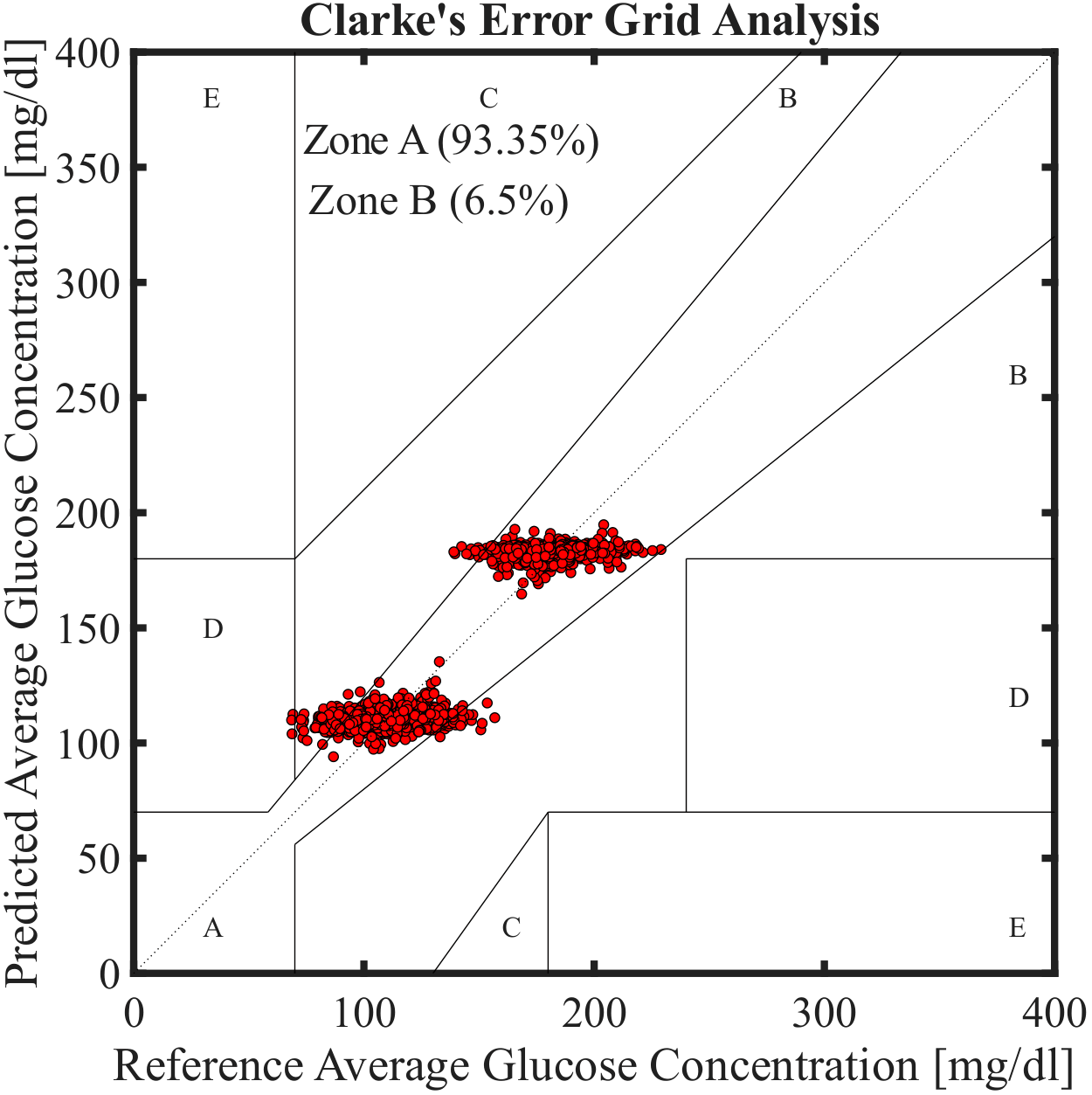}
	\caption{CEG analysis of training samples using MVPD4 model}
	\label{C4}
\end{figure}

The proposed DNN model has been examined with other computing model. It has been observed that the DNN model is comparatively precise for eAG and HbA1c prediction. The histogram plot has been represented to visualize the range in error for DNN model. Error histogram with 20 bins is presented for eAG prediction in Fig. \ref{hist}.
The correlation between predictors and output has also been represented during training, validation, and testing for eAG prediction. An optimized DNN computing model has been concluded for precise prediction. The fitting model represented is shown in Fig. \ref{nnfit}.
The samples are bifurcated for calibration and validation as per the standard ratio. The CEG analysis visualizes the error range in average glucose prediction and segregate the zone according to differences between the glucose values.
It has been confirmed that if the average glucose error in the predicted value is less than or equal to 20\% comparatively, then the system will be existed in zone A (clinically accurate) \cite{Clarke2005}. The predicted values in the range of 20\%-30\% error will be existed in zone B (Benign error).
The Clarke Error Grid (CEG) analysis has been done using multivariate polynomial model with degree 2, 3, and 4, and is represented by Fig. \ref{C2}, \ref{C3}, and \ref{C4}, respectively.

\begin{figure}[htbp]
	\centering
	\includegraphics[width=0.6\textwidth]{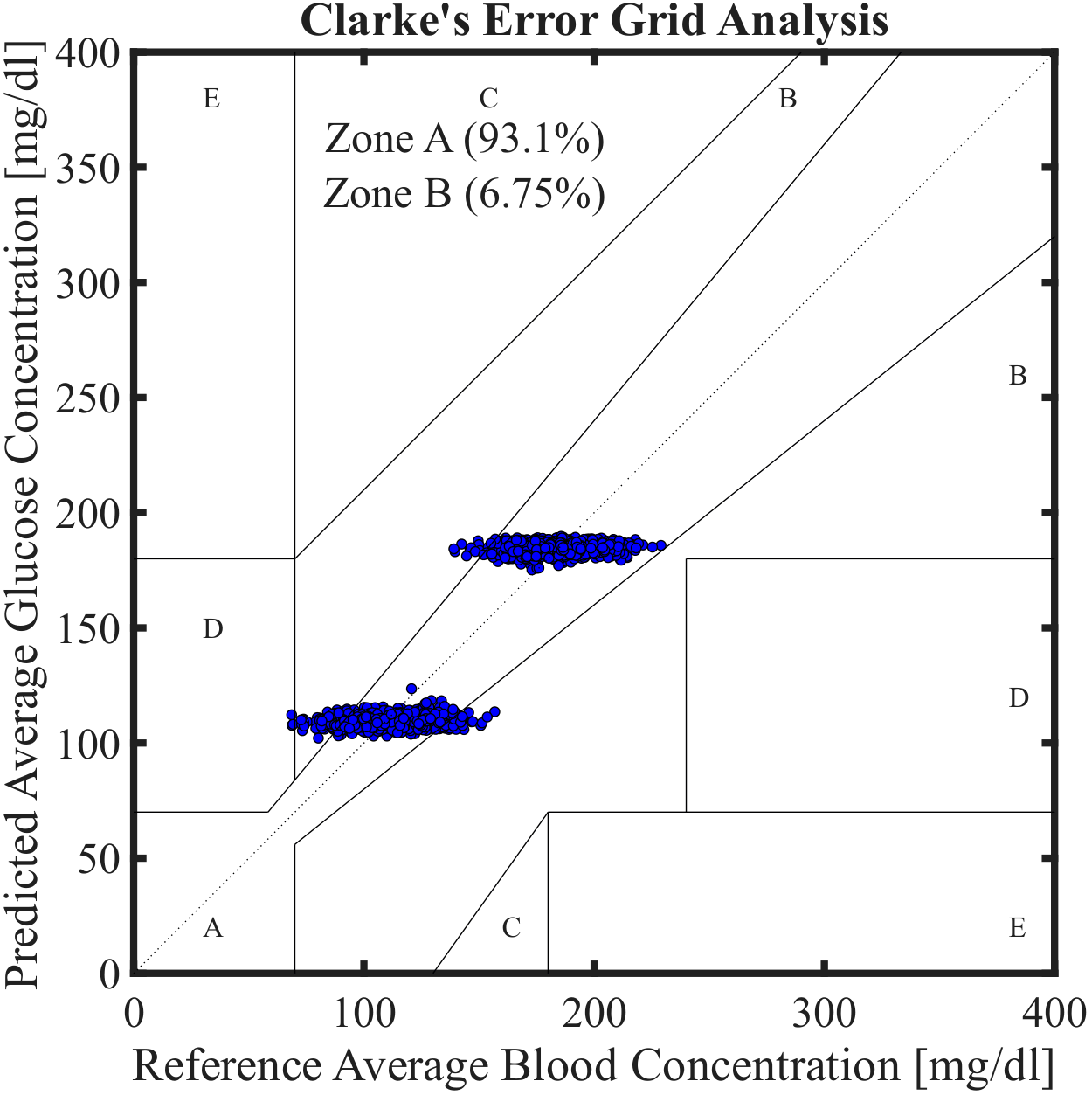}
	\caption{CEG analysis of training samples using DNN model}
	\label{1_DNN}
\end{figure}

\begin{figure}[htbp]
	\centering
	\includegraphics[width=0.6\textwidth]{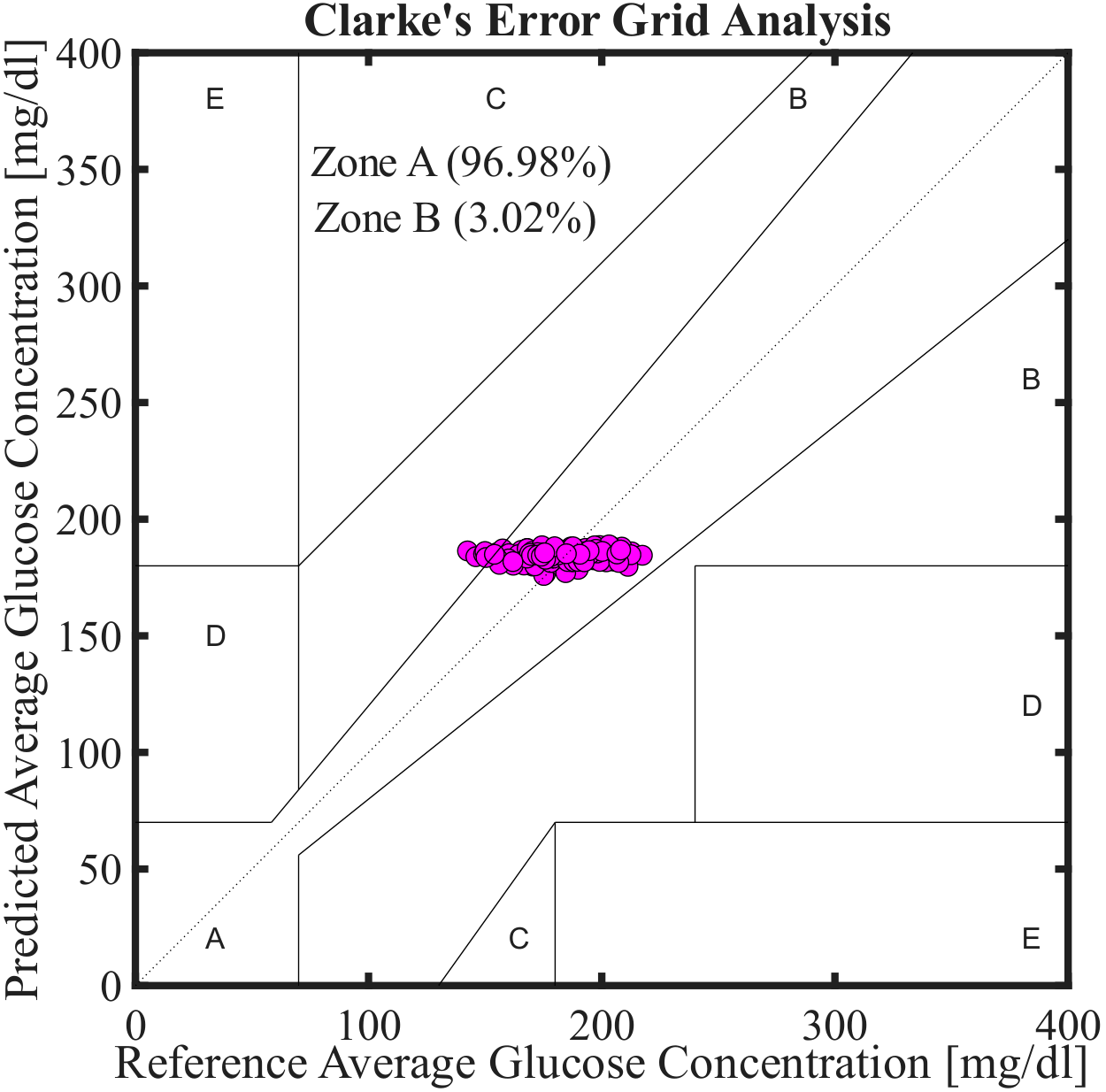}
	\caption{Clarke error grid analysis of randomly selected test samples using DNN model}
	\label{2_DNN}
\end{figure}

During CEG analysis, it has been concluded that 93.1\% of total observations have been found in zone A, and 6.75\% of total observations have been found in zone B during training and validation using optimized DNN model. During testing of 200 randomly picked samples, 96.98\% of total observations are found in zone A, and the remaining are found in zone B. The Clarke error grid (CEG) analysis of training and testing observations has been represented in Fig. \ref{1_DNN} and \ref{2_DNN}, respectively.
Comparative analysis of prior related work has been done to highlight the better results and technological advancement in the proposed work. The parametric analysis with technology has been represented in Table \ref{table_example}.
\begin{table}[htbp]
	\caption{Quantitative Analysis with Prior Works}
	\label{table_example}
	\setlength{\tabcolsep}{1.6pt}
	\centering
	\begin{tabular}{lllllll}
		\hline
		\textbf{Works}&{MAD}&mARD&Accuracy (\%)&Predictors&Samples&Measurement\\
		\hline	\hline
  Sridevi, et al.&-&-&92\% &PPG&45&Non-invasive\\
		\cite{sridevi2025noninvasive}&&&&signal&&\\
		\hline
		Mandal, et al.&-&-&90\%&in-vitro&45&Laboratory\\
		\cite{mandal2018vitro}&&&&sensing&&Test\\
		\hline
		Bent, et al.&-&-&91 \%&Body&26&Non-invasive\\
		\cite{bent2021non}&-&&&Parameters&&\\
		\hline
        \textbf{Proposed Work}&\textbf{0.37}&\textbf{0.048}&\textbf{4.67\% }&\textbf{Glucose}&\textbf{2000}&\textbf{Non-invasive}\\
  \textbf{iGLU 5.0}&&&\textbf{}&\textbf{Value and BP}&\textbf{}\\
		
		\hline
	\end{tabular}
\end{table}

\section{Conclusion and Future Direction}

The proposed iGLU 5.0 device is presented for non-invasive HbA1c measurement using blood glucose values and blood pressure. The basic sensing is based on optical detection. The proposed device utilized an optimized DNN model for precise eAG prediction. An average (training, validation, and testing) accuracy of 95\% has been observed using optimized DNN model for HbA1c prediction. The proposed device has been prepared for accurate HbA1c measurement, which will provide reliable support for rapid HbA1c prediction comparatively. The users will have the benefit of early diabetes prediction, and a medical representative can provide treatment on a prior basis for earlier control of glucose value fluctuations. The proposed device will have prime support for the patient at remote locations where the laboratory setup for measurement is not easily available. The proposed device will be helpful to segregate types of diabetes using precise HbA1c measurement. 

In the future plan, a non-invasive and reliable diabetes confirmation system will be proposed using dominant physiological parameters along with the HbA1c value. The system will have extensive support to provide the decision of diabetes confirmation to remotely located patients, where the medical representative and care centres are not easily available to provide the decision for further treatment. The future non-invasive system will be a low-cost, reliable and rapid decision-making system for smart healthcare. This will be the next step of technological advancement in the era of eMedicine. The user will get rapid treatment and continuous follow-up at remote locations.  

\section*{Acknowledgment}
The authors would like to express their sincere gratitude to the Dispensary, Malaviya National Institute of Technology and the System Level Design and Calibration Testing Lab. Special thanks are due to Nirma University for encouraging continued research with its support.

\bibliographystyle{IEEEtran}
\bibliography{arXiv_2026_iGLU5_HbA1c}

@article{mandali2023trends,
  title={Trends in quantification of HbA1c using electrochemical and point-of-care analyzers},
  author={Mandali, Pavan Kumar and Prabakaran, Amrish and Annadurai, Kasthuri and Krishnan, Uma Maheswari},
  journal={Sensors},
  volume={23},
  number={4},
  pages={1901},
  year={2023},
  publisher={MDPI}
}

@article{thapa2023label,
  title={Label-free electrochemical detection of glucose and glycated hemoglobin (HbA1c)},
  author={Thapa, Mukesh and Heo, Yun Seok},
  journal={Biosensors and Bioelectronics},
  volume={221},
  pages={114907},
  year={2023},
  publisher={Elsevier}
}

@article{ncd2023global,
  title={Global variation in diabetes diagnosis and prevalence based on fasting glucose and hemoglobin A1c},
  author={Majid Ezzati},
  journal={Nature medicine},
  volume={29},
  number={11},
  pages={2885--2901},
  year={2023},
  publisher={Nature Publishing Group US New York}
}

@article{hossain2021derivation,
  title={Derivation and validation of gray-box models to estimate noninvasive in-vivo percentage glycated hemoglobin using digital volume pulse waveform},
  author={Hossain, Shifat and Gupta, Shantanu Sen and Kwon, Tae-Ho and Kim, Ki-Doo},
  journal={Scientific reports},
  volume={11},
  number={1},
  pages={12169},
  year={2021},
  publisher={Nature Publishing Group UK London}
}

@article{kwon2022machine,
  title={Machine-learning-based noninvasive in vivo estimation of hba1c using photoplethysmography signals},
  author={Kwon, Tae-Ho and Kim, Ki-Doo},
  journal={Sensors},
  volume={22},
  number={8},
  pages={2963},
  year={2022},
  publisher={MDPI}
}

@article{arnardottir2023using,
  title={Using HbA1c measurements and the Finnish Diabetes Risk Score to identify undiagnosed individuals and those at risk of diabetes in primary care},
  author={Arnard{\'o}ttir, El{\'\i}n and Sigur{\dh}ard{\'o}ttir, {\'A}r{\'u}n K and Graue, Marit and Kolltveit, Beate-Christin Hope and Skinner, Timothy},
  journal={BMC Public Health},
  volume={23},
  number={1},
  pages={211},
  year={2023},
  publisher={Springer}
}

@article{joshi2020smart,
  title={Smart healthcare for diabetes: A COVID-19 perspective},
  author={Joshi, Amit M and Shukla, Urvashi P and Mohanty, Saraju P},
  journal={arXiv preprint arXiv:2008.11153},
  year={2020}
}

@article{agrawal2022machine,
  title={Machine learning models for non-invasive glucose measurement: towards diabetes management in smart healthcare},
  author={Agrawal, Harshita and Jain, Prateek and Joshi, Amit M},
  journal={Health and Technology},
  volume={12},
  number={5},
  pages={955--970},
  year={2022},
  publisher={Springer}
}

@InProceedings{Liu2016,
  author    = {Y. Liu and M. Xia and Z. Nie and J. Li and Y. Zeng and L. Wang},
  title     = {In vivo wearable non-invasive glucose monitoring based on dielectric spectroscopy},
  booktitle = {2016 IEEE 13th International Conference on Signal Processing (ICSP)},
  year      = {2016},
  pages     = {1388-1391},
  month     = {Nov},
  doi       = {10.1109/ICSP.2016.7878054},
  issn      = {2164-5221},
}

@Article{Song2015,
  author   = {K. Song and U. Ha and S. Park and J. Bae and H. J. Yoo},
  title    = {An Impedance and Multi-Wavelength Near-Infrared Spectroscopy IC for Non-Invasive Blood Glucose Estimation},
  journal  = {IEEE Journal of Solid-State Circuits},
  year     = {2015},
  volume   = {50},
  number   = {4},
  pages    = {1025-1037},
  month    = {April},
  issn     = {0018-9200},
  doi      = {10.1109/JSSC.2014.2384037},
}

@InProceedings{Sarangi2014,
  author    = {S. Sarangi and P. P. Pai and P. K. Sanki and S. Banerjee},
  title     = {Comparative Analysis of Golay Code Based Excitation and Coherent Averaging for Non-invasive Glucose Monitoring System},
  booktitle = {2014 IEEE 27th International Symposium on Computer-Based Medical Systems},
  year      = {2014},
  pages     = {485-486},
  month     = {May},
  doi       = {10.1109/CBMS.2014.102},
  issn      = {1063-7125},
}

@INPROCEEDINGS{Jain2024_1,
  author={Jain, Prateek and Joshi, Amit M. and Mohanty, Saraju P.},
  booktitle={2024 IEEE International Symposium on Smart Electronic Systems (iSES)}, 
  title={i{G}{L}{U} 4.1: An Intelligent Framework of Diabetes Prediction using Glucose-Insulin Values and Physiological Parameters}, 
  year={2024},
  volume={},
  number={},
  pages={315-320},
  doi={10.1109/iSES63344.2024.00073}}

@ARTICLE{MCE,
  author={Joshi, Amit M. and Jain, Prateek and Mohanty, Saraju P.},
  journal={IEEE Consumer Electronics Magazine}, 
  title={Everything You Wanted to Know About Continuous Glucose Monitoring}, 
  year={2021},
  volume={10},
  number={6},
  pages={61-66},
  doi={10.1109/MCE.2021.3073498}}

@Article{Pai2018,
  author    = {Pai, Praful P and De, Arijit and Banerjee, Swapna},
  title     = {Accuracy Enhancement for Noninvasive Glucose Estimation Using Dual-Wavelength Photoacoustic Measurements and Kernel-Based Calibration},
  journal   = {IEEE Transactions on Instrumentation and Measurement},
  year      = {2018},
  volume    = {67},
  number    = {1},
  pages     = {126--136},
  publisher = {IEEE},
}

@Article{Clarke2005,
  author    = {Clarke, William L},
  title     = {The original Clarke error grid analysis (EGA)},
  journal   = {Diabetes technology \& therapeutics},
  year      = {2005},
  volume    = {7},
  number    = {5},
  pages     = {776--779},
  publisher = {Mary Ann Liebert, Inc. 2 Madison Avenue Larchmont, NY 10538 USA},
}

@Article{jain2019precise,
  author    = {Jain, Prateek and Maddila, Ravi and Joshi, Amit M},
  title     = {{A precise non-invasive blood glucose measurement system using NIR spectroscopy and Huber’s regression model}},
  journal   = {Optical and Quantum Electronics},
  year      = {2019},
  volume    = {51},
  number    = {2},
  pages     = {51},
  publisher = {Springer},
}

@ARTICLE{iGLU2,
  author={Joshi, Amit M. and Jain, Prateek and Mohanty, Saraju P. and Agrawal, Navneet},
  journal={IEEE Transactions on Consumer Electronics}, 
  title={{iGLU 2.0: A New Wearable for Accurate Non-Invasive Continuous Serum Glucose Measurement in IoMT Framework}}, 
  year={2020},
  volume={66},
  number={4},
  pages={327-335},
  doi={10.1109/TCE.2020.3011966}}

@ARTICLE{iGLU3,
  author={Joshi, Amit M. and Jain, Prateek and Mohanty, Saraju P.},
  journal={IEEE Transactions on Consumer Electronics}, 
  title={{iGLU 3.0: A Secure Noninvasive Glucometer and Automatic Insulin Delivery System in IoMT}}, 
  year={2022},
  volume={68},
  number={1},
  pages={14-22},
  doi={10.1109/TCE.2022.3145055}}

@INPROCEEDINGS{9221132,
  author={Jain, Prateek and Joshi, Amit M. and Mohanty, Saraju P.},
  booktitle={2020 IEEE 6th World Forum on Internet of Things (WF-IoT)}, 
  title={{iGLU 1.1: Towards a Glucose-Insulin Model based Closed Loop IoMT Framework for Automatic Insulin Control of Diabetic Patients}}, 
  year={2020},
  volume={},
  number={},
  pages={1-6},
  doi={10.1109/WF-IoT48130.2020.9221132}}

@Article{Jain_IEEE-MCE_2020-Jan_iGLU1,
  author  = {Prateek Jain and Amit M. Joshi and Saraju P. Mohanty},
  title   = {{iGLU: An Intelligent Device for Accurate Non-Invasive Blood Glucose-Level Monitoring in Smart Healthcare}},
  journal = {IEEE Consumer Electronics Magazine},
  year    = {2020},
  volume  = {9},
  number  = {1},
  pages   = {35-42},
  month   = {January},
  doi     = {10.1109/MCE.2019.2940855},
}

@ARTICLE{new9431682,
	author={Althobaiti, Murad and Al-Naib, Ibraheem},
	journal={IEEE Photonics Journal}, 
	title={Optimization of Dual-Channel Near-Infrared Non-Invasive Glucose Level Measurement Sensors Based On Monte-Carlo Simulations}, 
	year={2021},
	volume={13},
	number={3},
	pages={1-9},
	doi={10.1109/JPHOT.2021.3079408}}

@ARTICLE{jainaccess,
  author={Jain, Prateek and Joshi, Amit M. and Mohanty, Saraju P. and Cenkeramaddi, Linga Reddy},
  journal={IEEE Access}, 
  title={Non-Invasive Glucose Measurement Technologies: Recent Advancements and Future Challenges}, 
  year={2024},
  volume={12},
  number={},
  pages={61907-61936},
  doi={10.1109/ACCESS.2024.3389819}}

@article{jain2024iglu,
  title={i{G}{L}{U} 4.0: Intelligent Non-invasive Glucose Measurement and Its Control with Physiological Parameters},
  author={Jain, Prateek and Joshi, Amit M and Mohanty, Saraju P},
  journal={SN Computer Science},
  volume={5},
  number={4},
  pages={368},
  year={2024},
  publisher={Springer}
}

@article{bent2021non,
  title={Non-invasive wearables for remote monitoring of HbA1c and glucose variability: proof of concept},
  author={Bent, Brinnae and Cho, Peter J and Wittmann, April and Thacker, Connie and Muppidi, Srikanth and Snyder, Michael and Crowley, Matthew J and Feinglos, Mark and Dunn, Jessilyn P},
  journal={BMJ Open Diabetes Research \& Care},
  volume={9},
  number={1},
  year={2021},
  publisher={American Diabetes Association}
}

@article{sridevi2025noninvasive,
  title={Noninvasive estimation of blood glucose and HbA1c using Quantum Machine Learning technique},
  author={Sridevi, Parama and Rabbani, Masud and Aziz, Md Hasanul and Upama, Paramita Basak and Mamun, Sayed Mashroor and Khan, Rumi Ahmed and Ahamed, Sheikh Iqbal},
  journal={Machine Learning with Applications},
  volume={19},
  pages={100626},
  year={2025},
  publisher={Elsevier}
}

@article{gough2023within,
  title={Within-subject variation of HbA1c: A systematic review and meta-analysis},
  author={Gough, Alex and Sitch, Alice and Ferris, Erica and Marshall, Tom},
  journal={PLoS One},
  volume={18},
  number={8},
  pages={e0289085},
  year={2023},
  publisher={Public Library of Science San Francisco, CA USA}
}

@article{beunen2022fasting,
  title={Fasting plasma glucose level to guide the need for an OGTT to screen for gestational diabetes mellitus},
  author={Beunen, Kaat and Neys, Astrid and Van Crombrugge, Paul and Moyson, Carolien and Verhaeghe, Johan and Vandeginste, Sofie and Verlaenen, Hilde and Vercammen, Chris and Maes, Toon and Dufraimont, Els and others},
  journal={Acta Diabetologica},
  volume={59},
  number={3},
  pages={381--394},
  year={2022},
  publisher={Springer}
}

@article{jamieson2023oral,
  title={Oral glucose tolerance test to diagnose gestational diabetes mellitus: Impact of variations in specimen handling},
  author={Jamieson, Emma L and Dimeski, Goce and Flatman, Robert and Hickman, Peter E and Jones, Graham Ross Dallas and Marley, Julia V and McIntyre, H David and McNeil, Alan R and Nolan, Christopher J and Potter, Julia M and others},
  journal={Clinical biochemistry},
  volume={115},
  pages={33--48},
  year={2023},
  publisher={Elsevier}
}

@article{gomez2022understanding,
  title={Understanding the clinical implications of differences between glucose management indicator and glycated haemoglobin},
  author={Gomez-Peralta, Fernando and Choudhary, Pratik and Cosson, Emmanuel and Irace, Concetta and Rami-Merhar, Birgit and Seibold, Alexander},
  journal={Diabetes, Obesity and Metabolism},
  volume={24},
  number={4},
  pages={599--608},
  year={2022},
  publisher={Wiley Online Library}
}

@article{dahal2023predicting,
  title={Predicting exotic annual grass abundance in rangelands of the western United States using various precipitation scenarios},
  author={Dahal, Devendra and Boyte, Stephen P and Oimoen, Michael J},
  journal={Rangeland Ecology \& Management},
  volume={90},
  pages={221--230},
  year={2023},
  publisher={Elsevier}
}

@article{kaliappan2024analyzing,
  title={Analyzing classification and feature selection strategies for diabetes prediction across diverse diabetes datasets},
  author={Kaliappan, Jayakumar and Saravana Kumar, IJ and Sundaravelan, S and Anesh, T and Rithik, RR and Singh, Yashbir and Vera-Garcia, Diana V and Himeur, Yassine and Mansoor, Wathiq and Atalla, Shadi and others},
  journal={Frontiers in Artificial Intelligence},
  volume={7},
  pages={1421751},
  year={2024},
  publisher={Frontiers Media SA}
}

@article{mandal2018vitro,
  title={An in-vitro optical sensor designed to estimate glycated hemoglobin levels},
  author={Mandal, Sanghamitra and Manasreh, MO},
  journal={Sensors},
  volume={18},
  number={4},
  pages={1084},
  year={2018},
  publisher={MDPI}
}

@INPROCEEDINGS{jain2025,
  author={Jain, Prateek and Joshi, Amit M. and Mohanty, Saraju P.},
  booktitle={2025 IEEE International Symposium on Smart Electronic Systems (iSES)}, 
  title={i{G}{L}{U} 4.2: A Non-Invasive Intelligent System for Diabetes Likelihood Prediction Using Glucose Values and Body Parameters}, 
  year={2025},
  volume={},
  number={},
  pages={60-65},
  doi={10.1109/iSES67504.2025.00022}}

@article{newkirubakaran2023antiallergic,
	title={Antiallergic Abdominal Belt for Human Glucose Level Measurement Using Microwave Active Sensor Antenna},
	author={Kirubakaran, SJ Jebasingh and Bennet, M Anto and Shanker, NR},
	journal={IEEE Sensors Journal},
	year={2023},
	publisher={IEEE}
}

@article{newmohammadi2023dual,
	title={Dual Frequency Microwave Resonator for Non-invasive detection of Aqueous Glucose},
	author={Mohammadi, Pejman and Mohammadi, Ali and Kara, Ali},
	journal={IEEE Sensors Journal},
	year={2023},
	publisher={IEEE}
}


\section*{Authors' Biographies}
\begin{minipage}[htbp]{\columnwidth}
	\begin{wrapfigure}{l}{1.0in}
		\vspace{-0.4cm}
		\includegraphics[width=1.0in,keepaspectratio]{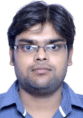}
		\vspace{-0.3cm}
	\end{wrapfigure}
	\noindent
	\textbf{Prateek Jain} (member IEEE) earned his B.E. degree in Electronics Engineering from Jiwaji University, India in 2010 and Master degree from ITM University Gwalior. He obtained PhD from Malaviya National Institute of Technology, Jaipur. He was awarded from MHRD fellowship during 2016-2020. He is currently Assistant Professor in Electronics \& Instrumentation Engg. Deptt., Institute of Technology, Nirma University, Ahmedabad (India). He was senior assistant professor in school of electronics engg. (SENSE), VIT AP University, AP since June 2020-Dec.2023. His current research interest includes Real-time system design, Biomedical Instrumentation, Low power VLSI design and Biomedical Systems. He is an author of 25 peer-reviewed publications. He is a regular reviewer of journals and 10 conferences. He was resource person in reputed universities for technical programs.
\end{minipage}

\vspace{0.8cm}

\begin{minipage}[htbp]{\columnwidth}
	\begin{wrapfigure}{l}{1.0in}
		\vspace{-0.4cm}
		\includegraphics[width=1.0in,keepaspectratio]{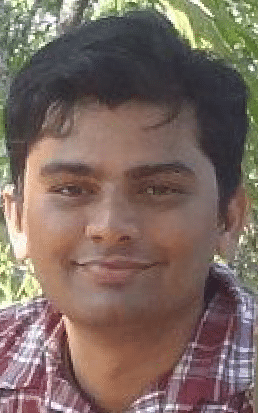}
		\vspace{-0.5cm}
	\end{wrapfigure}
	\noindent
	\textbf{Amit M. Joshi} (M'08) has completed his M.Tech (by research) in 2009 and obtained Doctoral of Philosophy degree (Ph.D) from National Institute of Technology, Surat in August 2015. He is currently an Assistant Professor at National Institute of Technology, Jaipur since July 2013. His area of specialization is Biomedical signal processing, Smart healthcare, VLSI DSP Systems and embedded system design. He has published six book chapters and also published 50+ research articles in peer reviewed international journals/conferences. He has served as a reviewer of technical journals such as IEEE Transactions, Springer, Elsevier and also served as Technical Programme Committee member for IEEE conferences. He also received UGC Travel fellowship, SERB DST Travel grant  and CSIR Travel fellowship to attend IEEE Conferences in VLSI and Embedded System. He has served session chair at various IEEE Conferences like TENCON -2016, iSES-2018, ICCIC-14. He has already supervised 18 M.Tech projects and 14 B.Tech projects in the field of VLSI and Embedded Systems and VLSI DSP systems. He is currently supervising six  Ph.D. students.
\end{minipage}

\vspace{0.5cm}

\begin{minipage}[htbp]{\columnwidth}
	\begin{wrapfigure}{l}{1.00in}
		\vspace{-0.3cm}
		\includegraphics[width=1.1in,keepaspectratio]{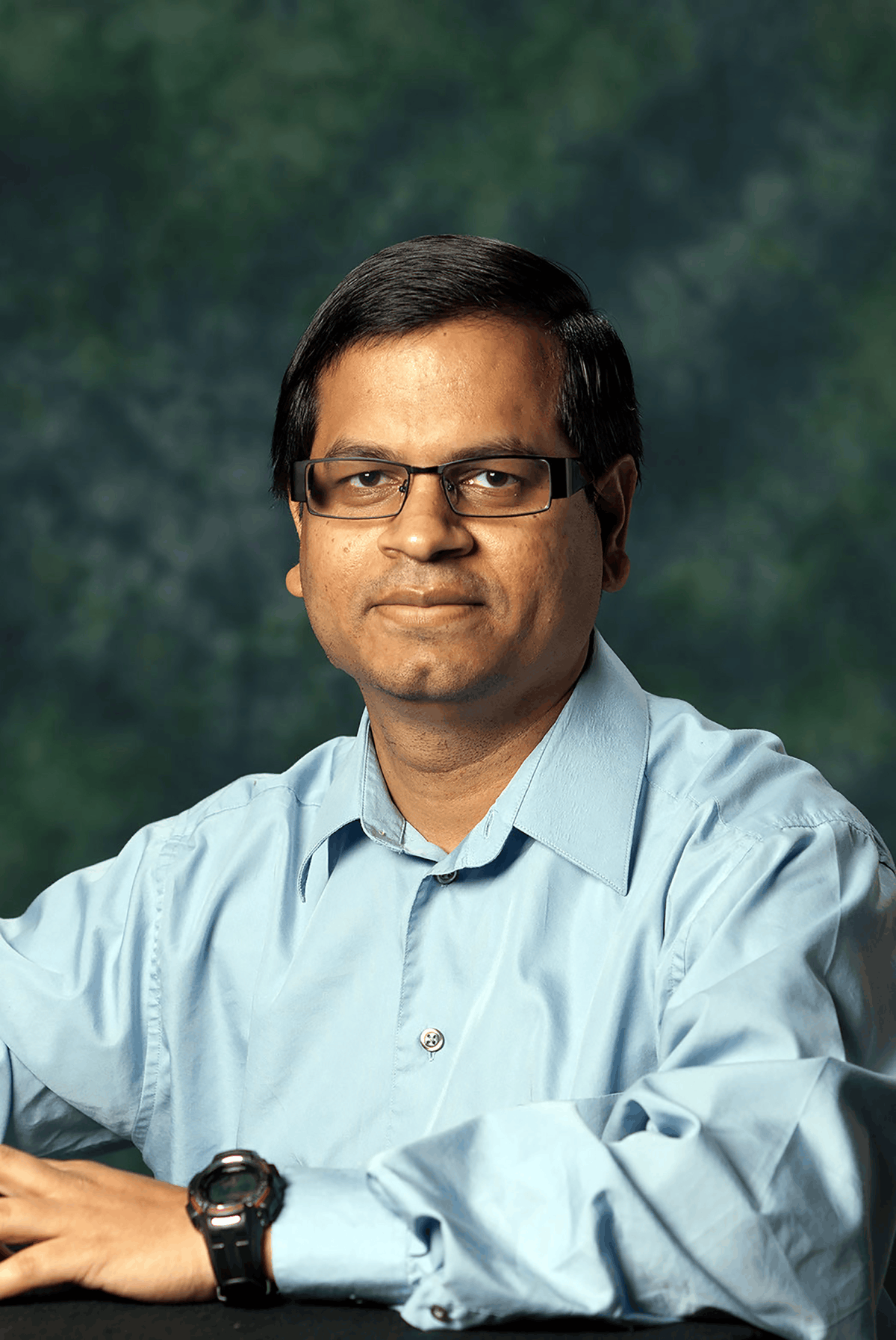}
		\vspace{-0.5cm}
	\end{wrapfigure}
	\noindent
	\textbf{Saraju P. Mohanty} (Senior Member, IEEE) received the bachelor’s degree (Honors) in electrical engineering from the Orissa University of Agriculture and Technology, Bhubaneswar, in 1995, the master’s degree in Systems Science and Automation from the Indian Institute of Science, Bengaluru, in 1999, and the Ph.D. degree in Computer Science and Engineering from the University of South Florida, Tampa, in 2003. He is a Professor with the University of North Texas. His research is in ``Smart Electronic Systems’’ which has been funded by National Science Foundations (NSF), Semiconductor Research Corporation (SRC), U.S. Air Force, NIDILRR, IUSSTF, and Mission Innovation. He has authored 600 research articles, 5 books, and 10 granted and pending patents. His Google Scholar h-index is 68 and i10-index is 338 with 19,000 citations. He is regarded as a visionary researcher on Smart Cities technology in which his research deals with security and energy aware, and AI/ML-integrated smart components. He introduced the Secure Digital Camera (SDC) in 2004 with built-in security features designed using Hardware Assisted Security (HAS) or Security by Design (SbD) principle. He is widely credited as the designer for the first digital watermarking chip in 2004 and first the low-power digital watermarking chip in 2006. He is a recipient of 21 best paper awards, Fulbright Specialist Award in 2021, IEEE Consumer Electronics Society Outstanding Service Award in 2020, the IEEE-CS-TCVLSI Distinguished Leadership Award in 2018, and the PROSE Award for Best Textbook in Physical Sciences and Mathematics category in 2016. He has delivered 33 keynotes and served on 15 panels at various International Conferences. He has been serving on the editorial board of several peer-reviewed international transactions/journals, including IEEE Transactions on Big Data (TBD), IEEE Transactions on Computer-Aided Design of Integrated Circuits and Systems (TCAD), IEEE Transactions on Consumer Electronics (TCE), and ACM Journal on Emerging Technologies in Computing Systems (JETC). He has been the Editor-in-Chief (EiC) of the IEEE Consumer Electronics Magazine (MCE) during 2016-2021. He served as the Chair of Technical Committee on Very Large Scale Integration (TCVLSI), IEEE Computer Society (IEEE-CS) during 2014-2018 and on the Board of Governors of the IEEE Consumer Electronics Society during 2019-2021. He serves on the steering, organizing, and program committees of several international conferences. He is the steering committee chair/vice-chair for the IEEE International Symposium on Smart Electronic Systems (IEEE-iSES), the IEEE-CS Symposium on VLSI (ISVLSI), and the OITS International Conference on Information Technology (OCIT). He has supervised 3 post-doctoral researchers, 21 Ph.D. dissertations, 29 M.S. theses, and 41 undergraduate projects.

\end{minipage}

\end{document}